\documentclass[%
 aip,
 jcp,
 amsmath,amssymb,
 reprint,%
]{revtex4-1}

\usepackage{graphicx}
\usepackage{dcolumn}
\usepackage{bm}

\usepackage[utf8]{inputenc}
\usepackage[T1]{fontenc}
\usepackage{mathptmx}
\usepackage{etoolbox}
\usepackage{physics}

\makeatletter
\def\@email#1#2{%
 \endgroup
 \patchcmd{\titleblock@produce}
  {\frontmatter@RRAPformat}
  {\frontmatter@RRAPformat{\produce@RRAP{*#1\href{mailto:#2}{#2}}}\frontmatter@RRAPformat}
  {}{}
}
\makeatother
\begin{document}

\preprint{AIP/123-QED}

\title{Quantum Markovian master equation in the  high-temperature limit}
\author{Ricardo C. Zamar}
\email{ricardo.zamar@unc.edu.ar}
\affiliation{ 
Facultad de Matemática, Astronomía, Física y Computación,
Universidad Nacional de Córdoba, Ciudad Universitaria, Córdoba, 5000,
Córdoba, Argentina
}
\author{J. Agustín Taboada}
 \email{agustin.taboada@mi.unc.edu.ar}
\affiliation{ 
Facultad de Matemática, Astronomía, Física y Computación,
Universidad Nacional de Córdoba, Ciudad Universitaria, Córdoba, 5000,
Córdoba, Argentina
}
\affiliation{Instituto Gaviola, CONICET, Córdoba, Argentina}

\begin{abstract}
We present a critical derivation of the high-temperature quantum Markovian master equation (HTME), examining its foundational assumptions, their quantum-mechanical implications, and its range of validity. Starting from the  Born-Markov master equation,   and combining the spin Hamiltonian eigenoperator formalism with a linear expansion in statistical coefficients---as the only assumption---we obtain a quantum dissipator that generalizes the Abragam-Redfield-Hubbard inhomogeneous master equation (ARH-IME). Our derivation naturally incorporates an additional term for non-thermal, high-order initial states, while reducing to ARH-IME for spin states evolving near thermal equilibrium (weak-order).	Through an alternative operator-based derivation of the HTME, we confirm these results and reveal a symmetrization condition for the spectral densities in the linear
    thermal regime. We rigorously analyze the internal consistency of both approaches and compare them with prior literature.  
To illustrate these findings, we study:  	
($i$) A canonical spin-½ system interacting with a bosonic bath, demonstrating first-principles symmetrization of spectral densities at high temperatures.
($ii$) Singlet-triplet conversion in a correlated two-spin system, where the ARH-IME fails, exposing the limitations of the weak-order hypothesis in strongly correlated regimes. Our results challenge the traditional boundaries of NMR spin-lattice relaxation theory and provide a refined framework for modeling open quantum systems beyond weak order.
\end{abstract}

\maketitle

\section{Introduction}

The study of open quantum system dynamics emerged between the 1950s and 1970s as irreversible processes began to be understood in terms of the quantum mechanical laws governing microscopic dynamics \cite{vkampen_1954,vhove_1957,fano_1957}.	
	In particular, pioneering theoretical work in Nuclear Magnetic Resonance (NMR)
	 led to the formulation of an equation that describes the irreversible spin dynamics through the reduced density operator. In this framework, a spin system $(S)$ interacts with an environment $(B)$ under an external magnetic field. The $S$-$B$ coupling	
	is described quantum mechanically, while the environment is regarded as a quantum thermal bath at constant temperature $T$, inducing Markovian dynamics in $S$. Those early developments introduced key concepts such as \textit{spin-lattice relaxation} and \textit{relaxation times}, describing the evolution of nuclear spins toward thermal equilibrium \cite{bpp48,wbloch53,redfield_1965,redfield_1957,bloch56,bloch57,abragam61,hubbard61,barbara_2020}.  
In parallel, advances in mathematical physics established the universal form that a dissipator must take to describe Markovian quantum processes--namely, a quantum dynamical semigroup. These findings, formalized in the Gorini-Kossakowski-Sudarshan-Lindblad (GKSL) theory \cite{gorini76, lindblad76, breuerpetrucc2001, rivas_2012, manzano2020, pascazio_2017} (and the references cited therein), provided a rigorous basis for describing relaxation in terms of the characteristic microscopic interactions for a large class of problems using the Born-Markov equation \cite{blum2011}.
	Indeed, it has been shown that when the conditions for the secular approximation are met, this equation adopts the universal structure predicted by quantum semigroup theory \cite{breuerpetrucc2001}. 
    
    A complete and consistent quantum treatment of spin-lattice relaxation requires, in principle, describing both spin and environment degrees of freedom quantum mechanically. Such a requirement poses significant challenges, particularly in systems where lattice fluctuations exhibit collective behavior, leading to correlated responses among spins within an ensemble \cite{bloch56,bloch57}.	
	This  obstacle was overcome in certain historical NMR problems, such
as Nuclear Quadrupole Resonance (NQR) relaxation in solids due to vibrations \cite{vankran} or NMR relaxation mediated by conduction electrons \cite{slichter}, where the spin-lattice interaction was treated fully quantum mechanically.
	Yet, in many cases, such as those where relaxation is dominated by
 complex molecular motions (e.g., rotational/translational diffusion in fluids or hindered rotations in solids) \cite{lukiloue,kowalewski2017}--a quantum description of the lattice degrees of freedom is hardly tractable. 	
	Therefore, a semiclassical approach treating spins quantum mechanically and the environment as a stochastic process was seen as a means to circumvent the difficulty.
	However, it was soon recognized that such an approach fails to explain the thermalization at finite temperatures, demanding a more complete theory.
    
	Historically, the strategy for achieving  an accessible and comprehensive description of spin-lattice relaxation was based on two conditions. The first, largely encountered in NMR, is the high-temperature regime, which means that the thermal coefficients associated with all spin transitions involved are small, i.e., $\beta_T\omega_i\equiv\hbar \omega_i/kT \ll 1, \forall i$. This condition was invoked to justify the linear approximation of the spin thermal equilibrium state. The second is the requirement that the spin state always be close to equilibrium. We refer to this assumption as the {\it weak-order} hypothesis. Based on these assumptions and starting from the Markovian master equation Abragam,  Redfield, and Hubbard independently reached  an inhomogeneous master equation (ARH-IME)  with the form \cite{abragam61,redfield_1965,hubbard61}
	\begin{equation}
		\label{inhomog_eq}
		\frac{d\rho_S}{dt} =   
		\Gamma_{HT} (\rho_S(t)-\rho_S^{eq}),
	\end{equation}
	where $\rho_S$  is the reduced spin state, $\rho_S^{eq}$ is the spin thermal equilibrium state and the  relaxation superoperator $\Gamma_{HT} $ has the same  double-commutator structure as the semiclassical master equation, which predicts that the system tends to equilibrium at infinite temperature regardless
of the actual finite bath temperature.    Unlike this approach, Eq.(\ref{inhomog_eq}) does not exhibit this drawback due to the presence of $\rho_S^{eq}$, which ensures relaxation to equilibrium at finite temperatures.
	This equation has been extensively applied to describe a  wide range of experiments \cite{ernst1990}. The key to its applicability lies in the assumption that, since thermal equilibrium is inherently guaranteed by Eq.(\ref{inhomog_eq}), it remains valid to model the environment stochastically \cite{Hilla-2025}.

Despite its empirical success, the theoretical status of Eq.(\ref{inhomog_eq}) remains debated. Questions persist  regarding the physical origin of the inhomogeneous term, the rigor of its derivations and whether it is merely an {\it ad-hoc} fix for the thermalization problem.

	In recent years there has been  renewed interest in revisiting the foundations of  NMR relaxation theory to disclose  its relationship with the GKSL theory \cite{bengs2020,rodin2021,barbara_2020,chruscinsky_2022}.
	In particular, the debate was reinvigorated by the work of Beng et al.\cite{bengs2020}, which studied the spin relaxation of a system of homonuclear proton pairs prepared in a high-order initial singlet state within the context  of long-lived states \cite{levitt2019}.	
	Equation  \eqref{inhomog_eq}, derived under the weak order assumption,  does not describe the slow polarization build-up when the system is  prepared in such non-thermal  state.	
	To analyze this phenomenon, the authors proposed an alternative approach that combines the Lindblad dissipator with  classical spectral densities suitably corrected to compensate for the loss of system-environment correlations necessary for thermal equilibrium  at finite temperatures. This procedure successfully described the slow polarization dynamics due to cross-relaxation between the Zeeman and singlet observables.

    Despite the long history of successful applications and recent theoretical progress achieved by incorporating methods from open quantum systems theory,  there still lingers a lack of consensus on the implications of the high-temperature limit as the main hypothesis for dictating spin system dynamics. 
    For instance, questions about the limitations imposed by the high temperature regime on the dynamics of highly correlated spin states as well on the nature of the system-environment coupling, remain. More fundamentally, the relationship between traditional NMR relaxation theory and Lindblad’s equation is still unclear.
Specifically, a new query emerges about whether Eq.(1) can be derived as a special case of the quantum Markovian framework.

To answer these profound questions, in this work, we derive a master equation starting from the quantum Markovian master equation, retaining only the high-temperature assumption while decoupling it from other common approximations as weak order or classical environment. Our goals are :

i) to obtain the limiting form of the quantum Markovian dissipator in the high-temperature regime.

    ii) to reveal the consequences that this regime has on the microscopic and coarse-grained timescales as well as on the quantum-mechanical character of the master equation.
	
    iii) to assess its applicability across diverse scenarios.
    
		We start from the Born-Markov equation, and employ the eigenoperator formalism to impose the secular approximation, obtaining a high-temperature-consistent master equation valid for arbitrary initial system states  (sections \ref{Sec.II} and \ref{Sec:3}). This general result is then applied to two scenarios: the spin-boson model and singlet-triplet conversion (Section \ref{Sec:4}).
	 In Sec. \ref{Sec.V} we reobtain the high-temperature master equation following the A. Abragam's operator formalism, but without invoking the weak order condition. While this yields the same formal result as in Sec. \ref{Sec.II}, it shows explicitly and in detail how the high temperature approximation modifies the structure of the Lindblad equation and its microscopic properties. 
		Finally, Sec.\ref{Sec.Concl} provides a discussion and summary of our findings.
	
\section{\label{Sec.II} The quantum Markovian master equation}
\subsection{The Born-Markov master equation}

Let us consider a composite quantum system described by the Hilbert space  $\mathcal{H_S}\otimes\mathcal{H_B}$, where $\mathcal{H_S}$ corresponds to the system of interest $S$ and $\mathcal{H_B}$ represents the environment or thermal bath $B$. $S$ and $B$ interact through the system-environment Hamiltonian $H_I$, and the entire system is described by
\begin{equation} 
\label{hamtot}
H = H_S + H_B + H_I\,, 
\end{equation}
where  $H_S$ characterizes the internal dynamics of $S$ (e.g. Zeeman interaction or chemical shift, spin-spin interactions such as dipole-dipole or J-coupling in a spin ensemble) and
Hamiltonian $H_B$ encodes  the environment's degrees of freedom (e.g. molecular reorientation in fluids or solids, order fluctuations in nematic liquid crystals, or a bosonic field). These operators satisfy the following general commutation rules:
\begin{equation}
\label{conmut}
[H_S,H_B]=0, \quad  [H_S,H_I]\neq 0 , \quad [H_I,H_B,]\neq 0\;.
\end{equation}

The irreversible dynamics of an open quantum system in the Markovian regime, can be described by the Born-Markov equation for  $\rho_S(t)$ in the interaction picture
\cite{blum2011,breuerpetrucc2001,rivas_2012}:
\begin{equation}
\label{bornmarkov}
\frac{d\rho_S}{dt} = -  \int_{0}^{\infty}d \tau\;\tr_B \left\lbrace \;
 [H_I(t),[H_I(t-\tau),  \rho_S(t) \otimes \rho_B ]]\;\right\rbrace  ,
\end{equation}
where tr$_B$ denotes partial trace over the environment degrees of freedom, and $\rho_S(t)\equiv \rm tr_B \rho (t)$ represents the reduced density operator of the system, obtained by tracing the density operator $\rho(t)$ of the compound system over the environment degrees of freedom.
On the other hand, $\rho_B$ is the density operator describing the thermal equilibrium state of the environment and the time dependence of $H_I$ arises from the transformation to the interaction picture $H_I(t)=e^{it(H_S+H_B)}H_Ie^{-it(H_S+H_B)}$. The validity of Eq. (\ref{bornmarkov}) hinges on the existence of two well-differentiated time scales: a microscopic one characterized by typical correlation times of the thermal bath, and the macroscopic time scale (coarse-grained scale) over which the system’s observables evolve. When the secular approximation  (or rotating wave approximation) holds, the complete positivity of $\rho_S(t)$ is ensured, and the master equation (\ref{bornmarkov}) adopts the standard GKSL form of a quantum Markovian  dissipator \cite{breuerpetrucc2001}.

Expanding the commutators  and rearranging the terms, Eq.(\ref{bornmarkov}) can be cast into a convenient form to identify the full-quantum contribution \cite{abragam61} (see Appendix \ref{AppendixA}):

\begin{widetext}
 \begin{equation}
 \label{bornmarkov_rearranged}
 \frac{d\rho_S}{dt} = - \frac{1}{2} \tr_B\left\{\int_{-\infty}^{\infty}d\tau\;
 [H_I(t),[H_I(t-\tau), \rho_S(t)   ]]\rho_B + \;  H_I(t) \rho_S(t) \int_{-\infty}^{\infty}d\tau\; [H_I(t-\tau),\rho_B]  -   \int_{-\infty}^{\infty}d\tau\; [H_I(t-\tau),\rho_B] \rho_S(t) 
  H_I(t)\right\} \;.
 \end{equation}
\end{widetext}
The replacement $\int_0^\infty \rightarrow \frac{1}{2}\int_{-\infty}^\infty$ is permissible since we have discarded the coherent contributions (Lamb shift)  \cite{breuerpetrucc2001}.
Notice that though we use a fully quantum description of the environment, the first term in (\ref{bornmarkov_rearranged}) exhibits the structure of a double commutator averaged over the environment degrees of freedom, resembling the semiclassical master equation. 
In the infinite temperature limit--or stochastic environment-- this term  remains finite, whereas the others vanish. For this reason, we refer to the remaining terms as full-quantum contributions.
Notice also that the thermal state $\rho_S^{eq}\propto e^{-\beta_T H_S}$ (with $\beta_T\equiv 1/k_BT$) is not a stationary state of the master equation when the full-quantum contributions are neglected, except at infinite temperature.  In ch.VIII of Abragam's book \cite{abragam61}, the last two terms in (\ref{bornmarkov_rearranged}) are interpreted as the source of correction that solves the failure of the semiclassical theory under the high-temperature and weak-order conditions.   

In the following, we demostrate that by treating the system-environment interaction quantum mechanically and assuming a high-temperature regime, a Markovian high-temperature master equation emerges as the limiting case of the Born-Markov equation.

\subsection{Master Equation in Terms of $\boldsymbol{H_S}$ eigenoperators. Secular Approximation}
To ensure  positivity of the spin dynamics governed by the quantum master equation, the secular aproximation must be  introduced. To further analyze Eq.(\ref{bornmarkov_rearranged}) under this approximation, we adopt a formalism based on the eigenoperators of $[H_S,\cdot]$.
Any system-environment Hamiltonian can be decomposed in the general form \cite{rivas_2012}
\begin{equation}
\label{hamiltoniano-interac}
H_I = \sum_{\alpha}  A_{\alpha}\otimes B_{\alpha} 
\;,
\end{equation}
where the operators $A_{\alpha}=A_{\alpha}^{\dagger}$ and  $B_{\alpha}=B_{\alpha}^{\dagger}$ act on the Hilbert spaces of the system and  the environment respectively. 
We write the operators $A_{\alpha}$ as a linear combination of the eigenoperators of $[H_S,\cdot]$, i.e., 
\begin{equation}
\label{autopexpand}
A_{\alpha} = \sum_{\omega}  A_{\alpha}(\omega),\quad A_{\alpha}(\omega) = \sum_{s - s'=\omega} {\cal P}(s) A_{\alpha}
{\cal P}(s')
\end{equation}
where ${\cal P}(s)$ is the projection operator onto the subspace corresponding to the eigenvalue $s$ of the Hamiltonian $H_S$, and $\omega= s-s'$ is a given energy difference \cite{breuerpetrucc2001}.
Substituting (\ref{autopexpand}) into (\ref{hamiltoniano-interac}),  $H_I$ in the interaction picture becomes
\begin{equation}
\label{hamiltonianos}
\begin{split}
    &H_I(t) = \sum_{\alpha,\omega'} e^{i\omega't} A^{\dagger}_{\alpha}(\omega')\otimes B_{\alpha}(t)\\
    &H_I(t-\tau) = \sum_{\beta,\omega} e^{-i\omega(t-\tau)} A_{\beta}(\omega)\otimes B_{\beta}(t-\tau)\;.
\end{split}
\end{equation}
Replacing Eq. (\ref{hamiltonianos}) in Eq.(\ref{bornmarkov_rearranged}), and using the secular approximation allows us to write respectively the double commutator  and the full-quantum terms as follows
\begin{subequations}
\begin{equation}
    \label{c-terms}
 \begin{split}
     -\frac{1}{2\hbar^2} \sum_{\alpha,\beta,\omega}&{\cal J}_{\alpha,\beta}(\omega)\{[A^\dagger_\alpha(\omega), [A_\beta(\omega), \rho_S(t) ]\\
     &\hspace{0,8cm}+(1-e^{-\beta_T \omega})[A_\beta(\omega), \rho_S(t)] 
 A^\dagger_\alpha(\omega) \} 
 \end{split}
\end{equation}
\begin{equation}
 \label{q-terms}
 \begin{split}
     \frac{1}{2\hbar^2}\sum_{\alpha,\beta,\omega}\mathcal{J}_{\alpha,\beta}(\omega)(1-&e^{-\beta_T\omega})([{A}_{\beta}(\omega)\rho_S(t),{A}_{\alpha}^{\dagger}(\omega)]\\
     &\hspace{0,8cm}-{A}_{\alpha}^{\dagger}(\omega)[\rho_S(t),{A}_{\beta}(\omega)]),
 \end{split}
\end{equation}
\end{subequations}

where the spectral densities are defined as 
\begin{equation}
\label{denpec_defin}
\begin{split}
    &{ \cal J}_{ \alpha \beta} (\omega)\equiv \int_{-\infty}^{\infty} \;d\tau \;
e^{i\omega \tau}  \;\tr\{ B_{\alpha}
B_{\beta}(-\tau)  \rho_B       \},\\
&\hspace{1.1cm} B_{\beta}(-\tau) = e^{-iH_B \tau} B_{\beta}\; e^{i H_B \tau}.
\end{split}
\end{equation}
The details of the derivations leading to Eqs. (\ref{c-terms}) and (\ref{q-terms}) can be found in Appendix \ref{AppendixB}.

From  
definition \eqref{denpec_defin}, we obtain the ``detailed-balance'' relations (see Appendix \ref{AppendixC})

\begin{subequations}
\begin{align}
\label{bce_detallado}
  {\cal J}_{ \alpha \beta} (\omega)&={ \cal K}_{\alpha\beta } (\omega)\; e^{\beta_T\omega}  \\ 
 \label{bce_detallado_b} {\cal J }_{\alpha\beta}(-\omega) &= {\cal J}_{\beta\alpha}(\omega) e^{-\beta_T\omega},
 \end{align}
\end{subequations}

where we introduced
\begin{subequations}
\begin{align}
    \label{Kdenpec_defin}
{ \cal K}_{ \alpha \beta} (\omega)&\equiv \int_{-\infty}^{\infty} \;d\tau \;
e^{i\omega \tau}  \;\tr\{ 
B_{\beta}(-\tau) B_{\alpha} \rho_B       \}\\
{ \cal J}_{ \beta \alpha} (\omega)&\equiv \int_{-\infty}^{\infty} \;d\tau \;
e^{i\omega \tau}  \;\tr\{ B_{\beta}
B_{\alpha}(-\tau)  \rho_B       \}.
\end{align}
\end{subequations}
Note that contribution (\ref{q-terms}) 
vanishes identically in both limits, $\beta_T \rightarrow 0$ and the classical reservoir approximation for the lattice. 
As highlighted in the previous section, we observe that (\ref{c-terms}) does not vanish at infinite temperature. Notably, it remains a non-zero term with the double commutator structure, resembling the form of the semiclassical master equation (see equation (42) in Chapter VIII of Abragam's book\cite{abragam61}). 

In conclusion, by gathering all the contributions and applying the secular approximation, the Markovian master equation takes the form
\begin{widetext}
    \begin{equation}
\label{ecmaes-eigenop}
\frac{d\hat\rho_S(t)}{dt} = - \frac{1}{2\hbar^2}\sum_{\alpha,\beta,\omega} {\cal J}_{\alpha\beta}(\omega)\{[A^\dagger_\alpha(\omega), [A_\beta(\omega), \rho_S(t) ]] -
(1-e^{-\beta_T \omega})[\rho_S(t)A_\beta(\omega),A^\dagger_\alpha(\omega) ] \}.
\end{equation}
\end{widetext}
A similar result was derived in the works of Goldman\cite{goldman2001}, Hubbard\cite{hubbard61} and Rodin et al.\cite{rodin2021}
 
Since the next section discusses the high temperature approximation, it is convenient to fully expose the dependence of Eq.(\ref{ecmaes-eigenop}) on the Boltzmann factors. Thus,  we separate the sum over $\omega$ into contributions from non-negative and negative frequencies.
  By using the identity  $A_{\alpha}(-\omega)=A_{\alpha}^{\dagger}(\omega)$ in the second sum, and suppressing the $\omega$ explicit dependency to short the notation, we can write  
\small
\begin{equation}
\label{eq:14}
    \begin{split}
        \Dot{\rho}_S&=-\frac{\hbar}{2}\sum_{\alpha\beta}\sum_{\omega\ge0}\mathcal{J}_{\alpha\beta}(\omega)([A_{\alpha}^{\dagger},[A_{\beta},\rho_S]]+(1-e^{-\beta_T\omega})[A_{\alpha}^{\dagger},\rho_SA_{\beta}])\\
        &-\frac{\hbar}{2}\sum_{\alpha\beta}\sum_{\omega
>0}\mathcal{J}_{\alpha\beta}(-\omega)([A_{\alpha},[A_{\beta}^{\dagger},\rho_S]]+(1-e^{\beta_T\omega})[A_{\alpha},\rho_SA_{\beta}^{\dagger}])
    \end{split}
\end{equation}
\normalsize
Focusing on the second term of Eq.(\ref{eq:14}), we now swap the dummy indices $\alpha,\beta$ (which preserves the sum):
\small
\begin{equation}
    \begin{split}
        &\sum_{\alpha\beta}\sum_{\omega
>0}\mathcal{J}_{\alpha\beta}(-\omega)([A_{\alpha},[A_{\beta}^{\dagger},\rho_S]]+(1-e^{\beta_T\omega})[A_{\alpha},\rho_SA_{\beta}^{\dagger}])\\
&=\sum_{\alpha\beta}\sum_{\omega
>0}\mathcal{J}_{\beta\alpha}(-\omega)([A_{\beta},[A_{\alpha
        }^{\dagger},\rho_S]]+(1-e^{\beta_T\omega})[A_{\beta},\rho_SA_{\alpha}^{\dagger}])
    \end{split}
\end{equation}
\normalsize
Then, using the balance relation $\mathcal{J}_{\beta\alpha}(-\omega)=e^{-\beta_T\omega}\mathcal{J}_{\alpha\beta}(\omega)$ (Eq.(\ref{bce_detallado_b})), the right-hand side becomes 
\small
\begin{equation}
\label{eq.16}
        \sum_{\alpha\beta}\sum_{\omega
>0}\mathcal{J}_{\alpha\beta}(\omega)(e^{-\beta_T\omega}[A_{\beta},[A_{\alpha
        }^{\dagger},\rho_S]]+(e^{-\beta_T\omega}-1)[A_{\beta},\rho_SA_{\alpha}^{\dagger}]).
\end{equation}
\normalsize
Now, we merge the transformed second term (Eq.(\ref{eq.16})) back into the original equation (\ref{eq:14}). The master equation now reads: 
\small
 \begin{equation}
    \begin{split}
        \Dot{\rho}_S=&-\frac{\hbar}{2}\sum_{\alpha\beta}\sum_{\omega\ge0}\mathcal{J}_{\alpha\beta}(\omega)\{[A_{\alpha}^{\dagger},[A_{\beta},\rho_S]]+e^{-\beta_T\omega}[A_{\beta},[A_{\alpha
        }^{\dagger},\rho_S]]\}\\
        &-\frac{\hbar}{2}\sum_{\alpha\beta}\sum_{\omega
>0}\mathcal{J}_{\alpha\beta}(\omega)\{(1-e^{-\beta_T\omega})[A_{\alpha}^{\dagger},\rho_SA_{\beta}]\\
&\hspace{2.5cm}+(e^{-\beta_T\omega}-1)[A_{\beta},\rho_SA_{\alpha}^{\dagger}]\},
    \end{split}
\end{equation}
\normalsize
This equation explicitly captures the full dependence of Eq.(\ref{ecmaes-eigenop}) on the Boltzmann factors derived from the detailed balance relations (\ref{bce_detallado}) and (\ref{bce_detallado_b}). Physically, this dependence arises from the commutation rules in Eq.(\ref{conmut}), which define the quantum properties of the system-environment interaction governing the microscopic scale.
\section{\label{Sec:3} High temperature approximation}
When the condition $\beta_T \omega\ll1$ is satisfied for all the system eigenfrequencies, it is reasonable
to approximate the master equation by keeping the linear term in the Taylor expansion of $e^{-\beta_T\omega}$ from Eq.(\ref{eq:14}) \cite{footnote3}. 
Thus,
\begin{equation}
\label{lineal}
    \begin{split}
        \Dot{\rho}_S&\approx-\frac{1}{2}\sum_{\alpha\beta}\sum_{\omega\ge0}\mathcal{J}_{\alpha\beta}(\omega)([A_{\alpha}^{\dagger},[A_{\beta},\rho_S]]+[A_{\beta},[A_{\alpha
        }^{\dagger},\rho_S]])\\
        &-\frac{1}{2}\sum_{\alpha\beta}\sum_{\omega>0}\mathcal{J}_{\alpha\beta}(\omega)\beta_T\omega([A_{\alpha}^{\dagger},\rho_SA_{\beta}]-[A_{\beta},\rho_SA_{\alpha}^{\dagger}]-[A_{\beta},[A_{\alpha
        }^{\dagger},\rho_S]]).
    \end{split}
\end{equation}
Proceeding with our analysis, we find it useful to write the  system state as 
 \begin{equation}
\label{desa-rho-llo}
\rho_S(t) = \frac{1}{Z} (\mathbb{I} + {\Delta \rho}_S (t) )\;,
\end{equation}
where $Z$ represents the number of degrees of freedom of the system. 
Equation (\ref{desa-rho-llo}) can be interpreted as a synthetic form of an expansion in  orthogonal operators,
\begin{equation}
\label{rhoexpan}
    {\rho}_S(t)=\frac{1}{Z}\left(\mathbb{I}+\sum_ic_i(t){O}_i\right),
\end{equation}
 where $\tr(O_i)=0$, $\tr({O}_i{O}^{\dagger}_j)=\delta_{ij}$, and $c_i(t)=\tr(\rho_S(t)O_i)$.
In this representation, the identity operator (representing complete disorder) is modified  
by a linear combination of orthogonal operators weighted by their respective mean values. These components represent the polarization and spin-spin correlations that can occur during the system's evolution \cite{QIP2024}.
Notably, some coefficients in (\ref{rhoexpan}) may be temperature-independent.
Substituting (\ref{desa-rho-llo}) into the second term of (\ref{lineal}) yields
\begin{equation}
\label{desglose}
    \begin{split}
        &\beta_T\omega([A_{\alpha}^{\dagger}, \frac{1}{Z} (\mathbb{I} + {\Delta \rho}_S (t) )A_{\beta}]-[A_{\beta}, \frac{1}{Z} (\mathbb{I} + {\Delta \rho}_S (t) )A_{\alpha}^{\dagger}]\\
        &-[A_{\beta},[A_{\alpha
        }^{\dagger}, \frac{1}{Z} (\mathbb{I} + {\Delta \rho}_S (t) )]]).
    \end{split}
\end{equation}
The operators $A_\beta(\omega)$, as eigen-operators of $[H_S,\cdot]$, satisfy the relation
\cite{breuerpetrucc2001}
\begin{equation}
\label{conmut_Hs}
\omega   A_\beta(\omega)= -[H_S, A_\beta(\omega) ].
\end{equation}
Thus, using (\ref{conmut_Hs}), and taking into account the linear expansion of $\rho_S^{eq}$ given by  
\begin{equation}
\label{rho_eq}
\rho_S^{eq}\simeq \frac{1}{Z}(\mathbb{I} - \beta_T H_S),
\end{equation}
we can rewrite each term in (\ref{desglose}) as follows:
\begin{widetext}
    \begin{equation}
\label{correction1}
    \begin{split}
        \frac{1}{Z}\beta_T\omega[A_{\alpha}^{\dagger},(I+\Delta\rho_S)A_{\beta}]=\frac{1}{Z}[A_{\alpha}^{\dagger},\beta_T\omega A_{\beta}]+\frac{1}{Z}[A_{\alpha}^{\dagger},\Delta\rho_S\beta_T\omega A_{\beta}]&=-\frac{\beta_T}{Z}[A_{\alpha}^{\dagger},[H_S, A_{\beta}]]-\frac{\beta_T}{Z}[A_{\alpha}^{\dagger},\Delta\rho_S[H_S,A_{\beta}]]\\
        &=[A_{\alpha}^{\dagger},[\rho_{eq},A_{\beta}]+[A_{\alpha}^{\dagger},\Delta\rho_S[\rho_{eq},A_{\beta}]]\\
        &=-[A_{\alpha}^{\dagger},[A_{\beta},\rho_{eq}]]+[A_{\alpha}^{\dagger},\Delta\rho_S[\rho_{eq},A_{\beta}]]
    \end{split}
\end{equation}
 \begin{equation}
   \begin{split}
        -\frac{\beta_T\omega}{Z}[A_{\beta},(I+\Delta\rho_S)A_{\alpha}^{\dagger}]=-\frac{\beta_T}{Z}[A_{\beta},\omega A_{\alpha}^{\dagger}]-\frac{\beta_T}{Z}[A_{\beta},\Delta\rho_S\omega A_{\alpha}^{\dagger}]&=-\frac{\beta_T}{Z}[A_{\beta},[H_S,A_{\alpha}^{\dagger}]]-\frac{\beta_T}{Z}[A_{\beta},\Delta\rho_S[H_S,A_{\alpha}^{\dagger}]]\\
        &=[A_{\beta},[\rho_{eq},A_{\alpha}^{\dagger}]]+[A_{\beta},\Delta\rho_S[\rho_{eq},A_{\alpha}^{\dagger}]]\\
        &=-[A_{\beta},[A_{\alpha}^{\dagger},\rho_{eq}]]+[A_{\beta},\Delta\rho_S[\rho_{eq},A_{\alpha}^{\dagger}]]
   \end{split}
\end{equation}
\begin{equation}
\label{correction3}
    -\frac{\beta_T}{Z}[A_{\beta},[A_{\alpha
        }^{\dagger},\rho_S]]=-\frac{\beta_T\omega}{Z}[\omega A_{\beta},[A_{\alpha
        }^{\dagger},\Delta\rho_S]]=\frac{\beta_T}{Z}[[H_S,A_{\beta}],[A_{\alpha
        }^{\dagger},\Delta\rho_S]]=-[[\rho_{eq},A_{\beta}],[A_{\alpha
        }^{\dagger},\Delta\rho_S]].
\end{equation}
\end{widetext}
Finally, gathering all the terms, we get
\begin{equation}
    \begin{split}
    \label{linenalizado}
        \Dot{\rho}_S&=-\frac{1}{2}\sum_{\alpha\beta}\sum_{\omega\ge0}\mathcal{J}_{\alpha\beta}(\omega)([A_{\alpha}^{\dagger},[A_{\beta},\rho_S-\rho_{eq}]]\\
        &\hspace{3cm}+[A_{\beta},[A_{\alpha
        }^{\dagger},\rho_S-\rho_{eq}]])\\
        &-\frac{1}{2}\sum_{\alpha\beta}\sum_{\omega>0}\mathcal{J}_{\alpha\beta}(\omega)([A_{\alpha}^{\dagger},\Delta\rho_S[\rho_{eq},A_{\beta}]]+[A_{\beta},\Delta\rho_S[\rho_{eq},A_{\alpha}^{\dagger}]]\\
        &\hspace{3cm}-[[\rho_{eq},A_{\beta}],[A_{\alpha
        }^{\dagger},\Delta\rho_S]]).
    \end{split}
\end{equation}
{In order to identify the ARH-IME equation within Eq.(\ref{linenalizado}), it is convenient to rewrite  
the contribution from the second and fourth terms in Eq.(\ref{linenalizado}) as }
\begin{equation}
    \begin{split}
 \label{eq.25}     &\sum_{\alpha\beta}\sum_{\omega>0}\mathcal{J}_{\alpha\beta}(\omega)([A_{\beta}(\omega),[A_{\alpha
        }^{\dagger}(\omega),\rho_S-\rho_{eq}]]\\
        &\hspace{0.6cm}+[A_{\beta}(\omega),\Delta\rho_S[\rho_{eq},A_{\alpha}(\omega)^{\dagger}]])\\
        &=\sum_{\alpha\beta}\sum_{\omega>0}\mathcal{J}_{\beta\alpha}(\omega)([A_{\alpha}(\omega),[A_{\beta
        }^{\dagger}(\omega),\rho_S-\rho_{eq}]]\\
        &\hspace{0.8cm}+[A_{\alpha}(\omega),\Delta\rho_S[\rho_{eq},A_{\beta}^{\dagger}(\omega)]])\\
        &=\sum_{\alpha\beta}\sum_{\omega>0}\mathcal{J}_{\alpha\beta}(-\omega)([A_{\alpha}^{\dagger}(-\omega),[A_{\beta
        }(-\omega),\rho_S-\rho_{eq}]]\\
        &\hspace{0.8cm}+[A_{\alpha}^{\dagger}(-\omega),\Delta\rho_S[\rho_{eq},A_{\beta}(-\omega)]]\\
        &=\sum_{\alpha\beta}\sum_{\omega<0}\mathcal{J}_{\alpha\beta}(\omega)([A_{\alpha}^{\dagger}(\omega),[A_{\beta
        }(\omega),\rho_S-\rho_{eq}]]\\&\hspace{0.8cm}+[A_{\alpha}^{\dagger}(\omega),\Delta\rho_S[\rho_{eq},A_{\beta}(\omega)]],
    \end{split}
\end{equation}
\normalsize
where, to maintain consistency with the linear approximation, we required the spectral densities in (\ref{linenalizado}) to be symmetric under the index exchange
	$\alpha \longleftrightarrow \beta$ and frequency 
	inversion $\omega\longleftrightarrow -\omega$, i.e., ${\cal J }_{\alpha\beta}(-\omega)= {\cal J}_{\beta\alpha}(\omega)$ (compare with eq. (\ref{bce_detallado_b})).

    Importantly, this symmetry does not reduce the environment to a classical regime.
The non-commutativity of the bath operators remains in the approximation, as the effects of the linear contribution from the detailed-balance relations in Eq.(\ref{ecmaes-eigenop}) are preserved.
 Moreover, conserving the statistical factor would introduce quadratic terms and higher in $\beta_T\omega$ already discarded. In any case, the information about the establishment of the thermal equilibrium at finite temperature is comprised as a consequence of the linear contribution from the quantum term in (\ref{ecmaes-eigenop}). 
In summary, the linear approximation gives the system-environment interaction a distinctive quantum character that brings about the symmetrization of spectral densities at high temperatures and changes in the form of the master equation that ensure the tendency of the system state toward thermal equilibrium at finite temperatures. Nevertheless, it does not imply that the system-environment interaction must be necessarily semi-classical. That said, a semi-classical description remains justifiable for weakly correlated spin systems, such as those in liquid-state NMR.

These concepts are concretely illustrated in Section \ref{Sec:4} by two physical examples: (i) a two level system coupled to a quantum thermal bath, and
(ii) a spin-1/2 pair interacting with a noise source. Furthermore, Section \ref{Sec.V} provides an independent verification via an argument derived from Abragam's formalism, ultimately confirming our conclusions about spectral densities through an alternative approach.

Using Eq.(\ref{eq.25}), the master equation  (\ref{linenalizado}) can be expressed as
\begin{equation}
\begin{split}
\label{ecmaes-eigen_aprox_b}
\Dot{\rho}_S&=-\frac {1}{2}\sum_{\alpha\beta,\omega} { J}_{\alpha\beta}(\omega) [A^\dagger_\alpha(\omega), [A_\beta(\omega),  { \rho} - \rho_S^{eq} ]] +\Gamma'(\rho_S)\\
&=\Gamma_{HT}(\rho_S-\rho_S^{eq})+\Gamma'(\rho_S)=:{\Gamma}(\rho_S)
\end{split}
\end{equation}
with
\begin{equation}
\label{correction}
   \begin{split}
       \Gamma'(\rho_S):&=\frac{1}{2}\sum_{\alpha\beta}\sum_{\omega}J_{\alpha\beta}(\omega)[{\Delta \rho_S}   
[\rho_S^{eq}, A_\beta(\omega)],A^\dagger_\alpha(\omega) ]\\
&-\frac{1}{2}\sum_{\alpha\beta}\sum_{\omega>0}J_{\alpha\beta}(\omega)[[\rho_S^{eq},A_{\alpha}(\omega)],[A_{\beta
        }^{\dagger}(\omega),\Delta\rho_S]]).
   \end{split}
\end{equation}
\normalsize
We refer to Eq.(\ref{ecmaes-eigen_aprox_b}) as the \textit{High-Temperature Master Equation} (HTME). In this equation, we introduced the symbol $ J_{\alpha\beta}(\omega)$ to indicate the transformation of the spectral densities $\mathcal{J}_{\alpha\beta}(\omega) \rightarrow J_{\alpha\beta}(\omega)$ into the symmetric forms in the high temperature limit.
This emphasizes the fact that we are describing a quantum relaxation process ($[H_B,H_I] \neq 0$), while maintaining a system-environment interaction consistent with $\langle B_\alpha B_{\beta}(-\tau) \rangle\approx\langle B_{\beta}(-\tau) B_\alpha   \rangle $. In other words, after linearizing the master equation, we still retain the quantum nature reflected in the factor $e^{-\beta_T\omega}$ from the detailed-balance relation. However, the quantum symmetric $J_{\alpha\beta}(\omega)$ reflects the actual ``quantumness'' of the system-environment interaction at high temperatures.

The linear approximation of the Boltzmann factor used to derive  Eq.(\ref{ecmaes-eigen_aprox_b}), led us to the specific forms in (\ref{desglose}) and (\ref{correction1})-(\ref{correction3}).
Accordingly, the HTME structurally consists of the ARH-IME plus a correction term that depends implicitly on $\beta_T \omega$ through $\rho_S^{eq}$.  Consequently, if the initial state of the system were close to  $\rho_S^{eq}$, then $\Delta\rho_S$ would be proportional to $\beta_T \omega$. 
For consistency with the linear approximation, any contribution from this term to Eq. (\ref{ecmaes-eigen_aprox_b}) should therefore be discarded. On the other hand, if the initial state is non-thermal, the deviation term $\Delta\rho_S$ may include temperature-independent terms which are thus relevant to the linearized dissipator. Therefore, the quantum master equation 
 comprises the ARH-IME plus a correction that can become significant for non-thermal states \cite{redfield_1965}.

Notice the first term of the HTME the steady state arises exclusively from the identity operator in (\ref{desa-rho-llo}),  recovering the well-known ARH-IME (\ref{inhomog_eq})  (under the secular approximation). On the other hand, $\Gamma'(\rho_S)$ originates  from polarizations and correlations.

Retracing our derivation, we recognize that Eq.(\ref{ecmaes-eigenop}) is indeed the Lindblad equation. The steps leading to (\ref{ecmaes-eigenop}) follow the standard ``microscopic derivation''  of the Lindblad form \cite{breuerpetrucc2001,   blum2011}. Therefore, the HTME emerges as the first-order term in the Taylor expansion of the Lindblad equation about $\beta_T\omega=0$. Furthermore, as $\rho_S(t)$ 
 approaches the equilibrium state $\rho_S^{eq}$, we 
observe that the dissipator ${\Gamma}(\rho_S(t))$  in Eq.(\ref{ecmaes-eigen_aprox_b}) 
converges to ${\Gamma}_{HT}(\rho_S(t)-\rho_S^{eq})$
 as in  Eq.(\ref{inhomog_eq}).
 This demonstrates that the Lindblad equation asymptotically reproduces the ARH-IME dynamics, explaining why the latter equation correctly describes systems near thermal equilibrium. 
 
It is instructive to compare the procedure presented in this section with the historical reasoning used to introduce the high-temperature condition within the weak-order assumption.  
As discussed in the seminal works of Redfield \cite{redfield_1965} (on page 20) and Hubbard \cite{hubbard61} (equation (106)), these authors exploited the fact that the thermal state $(\rho^{eq}_S)$ is  a stationary point of the Born-Markov equation. This allowed them to replace 
$\Gamma_{BM}(\rho_S)$ with $ \Gamma_{BM}(\rho_S - \rho^{eq}_S)$ in the Markovian master equation. Then, in the high-temperature regime, and assuming $\rho_S - \rho^{eq}_S$
becomes a small quantity (weak order), higher-order terms in $\beta_T\omega$
 were neglected. Unlike these earlier attempts, our formalism does not need to rely on the weak order condition to describe the spin system  dynamics in the high temperature regime.

\vspace{3mm}

\section{Application}
\label{Sec:4}
\subsection{Two level system}
{To better understand the constrains imposed by the high-temperature limit on the system-environment interplay and its impact on relations
 (\ref{bce_detallado}) and (\ref{bce_detallado_b}), we}
 consider a two-level system (TLS) coupled with an electric quantum field. This simple yet paradigmatic example from quantum optics has the advantage that all the relevant Hamiltonians are known, permitting a fully quantum mechanical treatment of the problem.
 This enables us to rigurously analyze the high-temperature approximation and sistematically examine how the system dynamics evolves as this limit is approached.

The {system-environment} interaction is characterized by dipole coupling, represented in the interaction picture as \cite{breuerpetrucc2001}:
\begin{equation}
    H_I(t)=\Vec{d}\hat{\sigma}_-e^{-i\omega t}+\Vec{d}^*\hat{\sigma}_+e^{i\omega t},\qquad H_S=\frac{\hbar\omega}{2}\sigma_z,
\end{equation}
where $\sigma_i$ with $i=x,y,z$ are the Pauli operators, 
$d\in\mathbb{C}^3$, $\hat{\sigma}_{\pm}\equiv\frac{1}{{{2}}}(\sigma_x\pm i\sigma_y)$.

The operators $\sigma_{\pm}$ are eigen-operators of $H_S$
\begin{equation}
    [\sigma_z,\sigma_+]=\sigma_+,\quad [\sigma_z,\sigma_-]=-\sigma_-
\end{equation}
so, $\Vec{A}(\omega)=\Vec{A}=\Vec{d}\sigma_-$, $\Vec{A}(-\omega)=\Vec{A}^{\dagger}=\Vec{d}^*\sigma_+$. 

The spectral density is given by
\begin{equation}
\label{eq:_optic_spectral_density}
    \mathcal{J}(\omega)=\frac{4\omega^3}{3\hbar c^3}(1+N(\omega)),\quad \mathcal{J}(-\omega)=\frac{4\omega^3}{3\hbar c^3}N(\omega),
\end{equation}
where $N(\omega)=\frac{1}{e^{\beta_T\omega}-1}$ is the photon occupation number. It can be readily seen that relation (\ref{bce_detallado_b}) is satisfied, and also that 
 if $\beta_T\omega\ll1\Longrightarrow e^{\beta_T\omega}-1\to0\Longrightarrow N\gg1$, and thus $\mathcal{J}(\omega)\approx \mathcal{J}(-\omega)$, that is to say, the spectral density becomes symmetric as the temperature increases.
 
The equation of motion (Lindblad equation), for the density operator written in the form (\ref{desa-rho-llo}) 
\begin{equation}
\label{eq:33}
    \rho_S(t)= \frac{1}{2}(I+\langle\vec{\sigma}\rangle(t)\cdot\vec{\sigma})=\begin{pmatrix}
        \frac{1}{2}(1+\langle\sigma_z\rangle(t)) & \langle\sigma_-\rangle(t)\\
           \langle\sigma_+\rangle(t) &\frac{1}{2}(1-\langle\sigma_z\rangle(t)),
        \end{pmatrix} 
\end{equation}
 is
\begin{widetext}
    \begin{equation}
\label{eq:Eq_Motion_Optic}
   \begin{split}
        \Dot{\rho}_S(t)&=\gamma_0(N+1)\left(\sigma_-\rho_S\sigma_+-\frac{1}{2}\sigma_+\sigma_-\rho_S-\frac{1}{2}\rho_S\sigma_+\sigma_-\right)+\gamma_0N\left(\sigma_+\rho_S\sigma_--\frac{1}{2}\sigma_-\sigma_+\rho_S-\frac{1}{2}\rho_S\sigma_-\sigma_+\right)\\
        &\equiv\gamma_0\left(\begin{array}{cc}
        \frac{N}{2}(1-\langle\sigma_z\rangle(t))-\frac{N+1}{2}(1+\langle\sigma_z\rangle(t)) & -(N+\frac{1}{2})\langle\sigma_-\rangle(t)  \\
        -(N+\frac{1}{2})\langle\sigma_+\rangle(t)    & -\frac{N}{2}(1-\langle\sigma_z\rangle(t))+\frac{N+1}{2}(1+\langle\sigma_z\rangle(t))
        \end{array}\right)
   \end{split}
\end{equation}
\end{widetext}
where $\gamma_0=\frac{4\omega^3|\Vec{d}|^2}{3\hbar c^3}$, and $\gamma=\gamma_0(2N+1)$, $N\equiv N(\omega)$. Taking the time derivative of Eq.(\ref{eq:33}) and equating it to Eq.(\ref{eq:Eq_Motion_Optic}), the Bloch equations are obtained
\begin{equation}
\label{eq:35}
    \begin{split}
       &\langle\Dot{\sigma}_{z}\rangle(t)=-{(2N+1)\gamma_0}\langle\sigma_z\rangle(t)-{\gamma_0}\\
       &\hspace{1cm}=-\gamma\left(\langle\sigma_z\rangle(t)+\frac{\gamma_0}{\gamma}\right)=-\gamma\left(\langle\sigma_z\rangle(t)-\langle\sigma_z\rangle_{eq}\right),\\
       &\langle\Dot{\sigma}_{\pm}\rangle=-\frac{1}{2}\gamma\langle\sigma_{\pm}\rangle(t),\quad \langle\sigma_z\rangle_{eq}:=-\frac{\gamma_0}{\gamma}
    \end{split}
\end{equation}
Accordingly, in the high-temperature limit, we have
\begin{equation}
\label{eq:36}
     \langle\Dot{\sigma}_{z}\rangle(t)\approx-2N\gamma_0\left(\langle\sigma_z\rangle(t)-\langle\sigma_z\rangle_{eq}\right),\quad \langle\Dot{\sigma}_{\pm}\rangle\approx-N\gamma_0\langle\sigma_{\pm}\rangle(t).
\end{equation}
On the other hand, using the ARH-IME, we obtain
\begin{equation}
     \begin{split}
         \Dot{\rho}&=-\frac{\gamma_0N}{2}([\sigma_+,[\sigma_-,\rho_S-\rho_S^{eq}]]+[\sigma_-,[\sigma_+,\rho_S-\rho_S^{eq}]])\\
         &=\left(\begin{array}{cc}
   -N\gamma_0(\langle\sigma_z\rangle(t)-\langle\sigma_z\rangle_{eq}) & -N\gamma_0\langle\sigma_-\rangle(t) \\
   -N\gamma_0\langle\sigma_+\rangle(t)    &  N\gamma_0(\langle\sigma_z\rangle(t)-\langle\sigma_z\rangle_{eq})
    \end{array}\right).
     \end{split}
\end{equation}
Thus
\begin{equation}
\label{eq:AR_sol}
    \langle\Dot{\sigma}_z\rangle(t)=-2N\gamma_0(\langle\sigma_z\rangle(t)-\langle\sigma_z\rangle_{eq}),\quad  \langle\Dot{\sigma}_{\pm}\rangle(t)=-N\gamma_0\langle\sigma_{\pm}\rangle(t).
\end{equation}
Finally, using the HTME
\begin{equation}
    \begin{split}
        \Dot{\rho}&=-\frac{N\gamma_0}{2}([\sigma_+,[\sigma_-,\rho_S]]+[\sigma_-,[\sigma_+,\rho_S]])\\
        &+\frac{N\gamma_0}{2}([\sigma_-,[\sigma_+,\rho_{eq}]]+[\sigma_+,[\sigma_-,\rho_{eq}]])\\
        &+\frac{N\gamma_0}{2}\frac{\beta_T\omega}{2}([\sigma_-,[\sigma_+,\Delta\rho_S]]+[\sigma_-,\Delta\rho_S\sigma_+]-[\sigma_+,\Delta\rho_S\sigma_-])\\
        &=N\gamma_0\begin{pmatrix}
\langle\sigma_z\rangle_{eq}+(\frac{\beta_T\omega}{2}-1)\langle\sigma_z\rangle(t) & (\frac{\beta_T\omega}{2}-1)\langle\sigma_{-}\rangle(t)\\
(\frac{\beta_T\omega}{2}-1)\langle\sigma_{+}\rangle(t) & (1-\frac{\beta_T\omega}{2})\langle\sigma_z\rangle(t)-\langle\sigma_z\rangle_{eq}
        \end{pmatrix}
    \end{split}
\end{equation}
we find
\begin{equation}
\label{eq:HTME_sol}
    \begin{split}
         &\langle\Dot{\sigma}_z\rangle(t)=2N\gamma_0(\langle\sigma_z\rangle(t)(\frac{\beta_T\omega}{2}-1)+\langle\sigma_z\rangle_{eq})\\
         &\hspace{1cm}\approx 2N\gamma_0(\langle\sigma_z\rangle_{eq}-\langle\sigma_z\rangle(t))\\
         &\langle\Dot{\sigma}_-\rangle(t)=N\gamma_0(\frac{\beta_T\omega}{2}-1)\langle\sigma_-\rangle(t)\approx -N\gamma_0\langle\sigma_-\rangle(t).
    \end{split}
\end{equation}
By comparing (\ref{eq:36}) and (\ref{eq:AR_sol}), 
we observe that the dynamics governed by Eq.(\ref{eq:Eq_Motion_Optic})  naturally converge to that governed by the ARH-IME in the high-temperature limit. Notice also that within this limit, the spectral density becomes
symmetric: the original spectral density (\ref{eq:_optic_spectral_density}), describing both spontaneous and stimulated transitions, transforms into one associated solely with stimulated transitions (absorption and emission). The corresponding transition probabilities
become equal—consistent with semiclassical models—whereas spontaneous emission lacks a semiclassical counterpart  \cite{GerryKnight2023}. As a consequence, the HTME cannot address relaxation due to vacuum fluctuations. Nevertheless, the high-temperature master equation adequately describes finite-temperature
thermalization.
In conclusion, Lindblad’s and ARH’s equations produce identical dynamics at high temperatures, with a symmetric spectral density and independently of the TLS initial condition. Moreover, the contribution $\Gamma'(\rho_S)$ from Eq.(\ref{correction}) turns negligible in this scenario, suggesting that its relevance depends not only on the initial state preparation but also on the system interactions.

Although simple, this example illustrates that spectral density symmetrization can occur despite the environment’s quantum nature, showing that a quantum description of the environment can be compatible with symmetric spectral densities in the high-temperature regime.

Finally, we note a parallel with Hebel and Slichter approach \cite{slichter} to nuclear spin relaxation in metals due to the coupling with conduction electrons. There, the spin transition rates ($W_{nm}$) symmetrize under a high-temperature approximation, replacing the detailed balance condition $W_{mn}=W_{nm}e^{(E_m-E_n)\beta_T}$. This quantumness loss in the transition probabilities (microscopic scale) aligns with the linear-temperature regime of the master equation (coarse-grained scale). The authors derive a spin relaxation rate from a linearized Pauli equation in $\beta_T \omega$, discarding higher-order terms in $\beta(E_m-E_n)$ in the rate equation
\begin{equation}
    \frac{dp_n}{dt}=\sum_m(p_mW_{mn}-p_nW_{nm}),
\end{equation}
yielding a first-order differential equation for the spin temperature.

\subsection{Singlet-Triplet Conversion}

We now apply  equation  Eq. (\ref{ecmaes-eigen_aprox_b}) to study the dynamics of a spin-1/2 pair coupled to a random magnetic field, where the spin system is prepared in a non-thermal initial state.
 This problem was addressed by Rodin et al.\cite{rodin2021}, and Bengs et al.\cite{bengs2020}, which suggested that the ARH-IME fails to correctly describe the dynamics when the system is initialized with saturated singlet order.
  In particular, Bengs et al\cite{bengs2020}, tackled the issue using Lindblad's equation, however, due to the complexity of a full quantum-mechanical treatment of the environment
  they replaced the quantum spectral densities with their classical counterparts, weighted by the factor $e^{\beta_T\omega/2}$ to ensure the detailed balance condition (see references \cite{redfield_1965, goldman1988}). This approach predicts a bi-exponential decay of the longitudinal magnetization due to spin-lattice cross relaxation, a feature absent in the ARH-IME single-exponential prediction, which explains the slow recovery of the magnetization with a time constant larger than $T_1$.

In this section, we demonstrate that the HTME  (\ref{ecmaes-eigen_aprox_b}) reproduces the same bi-exponential decay for this model. The interaction Hamiltonian is given by:
\begin{equation}
\label{Interaccion_LB}
    H_I=\sum_{q=-1}^1\sum_{i=1}^2B_{ran}^{(i)}(t)T_q^{(i)}
\end{equation}
where $T_q^{(i)}$ are spherical tensors of rank 1 corresponding to the $i$-th spin, and $B_{ran}^{(i)}(t)$ is a random field at site $i$. 

Inserting (\ref{Interaccion_LB}) into (\ref{ecmaes-eigen_aprox_b}), we get
\begin{equation}
\label{gamma_ran}
  \begin{split}
      \Dot{\rho_S}=\Gamma_{ran}(\rho_S):&=-\dfrac{\omega_{ran}^2}{2}\sum_{ij}\kappa_{ij}\sum_{q=-1}^{1}J(q\omega_0)([(T_q^i)^{\dagger},[T_q^j,\rho_S-\rho_S^{eq}]]\\
      &\hspace{2.5cm}-[\Delta\rho_S[\rho_S^{eq},T_q^i],(T_q^j)^{\dagger}])\\
      &+\sum_{ij}J(\omega_0)\kappa_{ij}[[\rho_{eq},T_1^i],[(T_{1
        }^j)^{\dagger},\Delta\rho_S]])
  \end{split}
\end{equation}
The \textit{classical} spectral densities
\begin{equation}
    J_{ij}(q\omega_0) := \int_{\mathbb{R}} e^{-iq\omega_0\tau}  d\tau \langle B_{ran}^{(i)}(0) B_{ran}^{(j)}(\tau) \rangle
\end{equation}
are assumed to be\cite{bengs2020} $J_{ij}(q\omega_0) = \kappa_{ij} J(q\omega_0)$, where $J(q\omega_0) = \frac{2\tau_{ran}}{1 + (\omega_0 \tau_{ran})^2}$. Here, $\kappa_{ij}$ quantifies the correlation between the fields at sites $i$ and $j$ ($\kappa_{ii} = 1$ and $\kappa_{ij} = \kappa_{ji}$), and $\tau_{ran}$ represents the characteristic time of the autocorrelation functions.
When $\kappa_{ij}=0$, the system-bath interaction model coincides with that of Wangsness and Bloch \cite{wbloch53}, and Bloembergen, Purcell and Pound \cite{bpp48}, namely individual spins interacting with uncorrelated reservoirs. Finally, the term $\omega_{ran}$ is the root-mean-square amplitude of the local fluctuations.
We now show that the presence of the correction term in (\ref{ecmaes-eigen_aprox_b}) is the source of the coupling between these observables. Before proceeding, and to facilitate comparison with the results of references \cite{rodin2021,bengs2020}, we make a key assumption:
the state of the systems is assumed to evolve within the diagonal subspace
spanned by the operator set
$B=\{I,I_z,T_{20},\Vec{I}_1\cdot\Vec{I}_2\}$, where $I$ is the identity operator, $I_z$ is the $z$ component of the total spin angular momentum, $T_{20}$ is the spherical tensor associated to secular part of the dipolar Hamiltonian and $\Vec{I}_1\cdot\Vec{I}_2$, represents the excess of singlet order\cite{keller88}. 
Consistent with this assumption --and in alignment with Eqs. (\ref{desa-rho-llo}) and (\ref{rhoexpan}) for the density operator-- we write
\begin{equation}
\label{rho_sol}
    \rho_S(t)=\frac{1}{4}(I+c_Z(t)I_Z+c_S(t)\Vec{I}_1\cdot\Vec{I}_2+c_D(t)T_{20})\equiv\dfrac{1}{4}(I+\Delta\rho_S(t)),
\end{equation}
where the operator $\Delta\rho_S(t)=c_1(t)I_z+c_2(t)\Vec{I}_1\cdot\Vec{I}_2+c_3(t)T_{20}$ is recognized as the deviation from the total disorder. In addition, we note that $c_Z(t)\propto\langle I_z\rangle(t)$, $c_S(t)\propto\langle\Vec{I}_1\cdot\Vec{I}_2\rangle(t)$ y $c_D(t)\propto\langle T_{20}\rangle(t)$. Consequently, we can write an equation for the coefficients $c_i(t)$ (i.e., for the expectation values). Defining $\vec{x}(t):=(1,c_S(t),c_D(t),c_Z(t))$, we get
\begin{equation}
\label{matrix_G}
\Dot{\vec{x}}(t)=\Gamma\vec{x}(t),\qquad\Gamma=\begin{pmatrix}
0 & 0 & 0 & 0\\ 
0 & \sigma_{SS} & \sigma_{SD} & \sigma_{SZ}\\ 
0 & \sigma_{DS} & \sigma_{DD} & \sigma_{DZ}\\ 
\sigma_{ZI} & \sigma_{ZS} & \sigma_{ZD} & \sigma_{ZZ}
\end{pmatrix}
\end{equation}
where the matrix elements in (\ref{matrix_G}) are defined as
\begin{equation*}
    \sigma_{ij}=\frac{\tr(\mathcal{O}^{\dagger}_i\Gamma_{ran}(\mathcal{O}_j))}{\sqrt{\tr(\mathcal{O}^{\dagger}_i\mathcal{O}_i)\tr(\mathcal{O}^{\dagger}_j\mathcal{O}_j)}},\quad\mathcal{O}_j=\{I,I_Z,\Vec{I}_1\cdot\Vec{I}_2,T_{20}\} 
\end{equation*}
For example, for the coefficients $\sigma_{ZS}$ and $\sigma_{SZ}$, we have
\begin{equation}
    \begin{split}
        &\sigma_{S,Z}:=\frac{\tr((\Vec{I}_1\cdot\Vec{I}_2)\Gamma_{ran}(I_z))}{\sqrt{\tr((\Vec{I}_1\cdot\Vec{I}_2)^2)\tr(I_z^2)}}\\
        &=\sqrt{\dfrac{2}{3}}\dfrac{\omega_{ran}^2}{2}\beta_T\omega_0\sum_{ij}\kappa_{ij}\sum_{q=-1}^{1}J(q\omega_0)\tr(\Vec{I}_1\cdot\Vec{I}_2[I_z[I_z,T_q^i],(T_q^j)^{\dagger}])\\
        &=\dfrac{1}{\sqrt{6}}{\omega_{ran}^2}\beta_T\omega_0\left[\dfrac{J(\omega_0)+J(-\omega_0)}{2}-\dfrac{\kappa_{12}}{2}(J(\omega_0)+J(-\omega_0))\right]\\
        &=\dfrac{1}{\sqrt{6}}{\omega_{ran}^2}\beta_T\omega_0J(\omega_0)(1-\kappa_{12})\approx\dfrac{1}{\sqrt{6}}\beta_T\omega_02\omega_{ran}^2\tau_{ran}(1-\kappa_{12})
    \end{split}
\end{equation}
\begin{equation}
    \begin{split}
        &\sigma_{Z,S}=\frac{\tr(I_z\Gamma_{ran}(\Vec{I}_1\cdot\Vec{I}_2))}{\sqrt{\tr((\Vec{I}_1\cdot\Vec{I}_2)^2)\tr(I_z^2)}}\\
        &=\sqrt{\dfrac{2}{3}}\dfrac{\omega_{ran}^2}{2}\beta_T\omega_0\sum_{ij}\kappa_{ij}\sum_{q=-1}^{1}J(q\omega_0)\tr(I_z[\Vec{I}_1\cdot\Vec{I}_2[I_z,T_q^i],(T_q^j)^{\dagger}])\\
        &=\sqrt{\dfrac{2}{3}}\dfrac{\omega_{ran}^2}{2}\beta_T\omega_0\kappa_{12}\left[\dfrac{J(\omega_0)+J(-\omega_0)}{2}\right]\\
        &=\dfrac{1}{\sqrt{6}}{\omega_{ran}^2}\beta_T\omega_0J(\omega_0)\kappa_{12}\approx\dfrac{1}{\sqrt{6}}\beta_T\omega_0\tau_{ran}2\omega_{ran}^2\tau_{ran}\kappa_{12},
    \end{split}
\end{equation}
where the narrowing limit $\tau_{ran}\omega_0\ll1$ is considered.

We recall from Sec.\ref{Sec.II} that the spectral densities in the HTME are symmetric (due to the diminished quantum effects under high-temperature conditions). Despite this, the overall structure of Eq.(\ref{ecmaes-eigen_aprox_b}) inherently provides the correct tendency towards equilibrium. Therefore, if the quantum nature of the problem under study permits the environment to be modeled as a stochastic bath, it becomes appropriate to use classical spectral densities, as we do here, for computing relaxation times and coupling coefficients.

The first and last terms in (\ref{gamma_ran}) do not contribute to the coupling coefficients, but define the relaxation times $T_1=\sigma_{ZZ}^{-1},$ $T_s=\sigma_{SS}^{-1}$ and $T_{1D}=\sigma_{DD}^{-1}$. Nevertheless, we neglect the contribution from the last term since it is proportional to $\beta_T\omega$ and thus negligible.

The remaining coefficients are given by
\begin{equation}
   \begin{split}
       &\sigma_{SS}=-T_s^{-1},\quad  \sigma_{SD}=\sigma_{DS}\approx0,\quad\sigma_{SZ}=\frac{R_{1r}}{\sqrt{6}}(1-\kappa)\beta_{T}\omega_0\\
       &\sigma_{DZ}=\frac{\sqrt{3}}{6}{R_{1r}}(\kappa+2){\beta_{T}\omega_0},\quad R_{1r}=2\omega_{ran}^2\tau_{ran}\\
       & \sigma_{ZI}=-\frac{1}{\sqrt{2}}R_{1r}\beta_{T}\omega_0,\quad\sigma_{ZD}=\frac{R_{1r}\kappa\beta_{T}\omega_0}{2\sqrt{3}}\\
       &\sigma_{DD}=-{R_{1r}}(\kappa+2),\quad\sigma_{ZS}=\frac{1}{\sqrt{6}}R_{1r}\kappa\beta_{T}\omega_0,\quad\sigma_{ZZ}=-T_1^{-1}
   \end{split}
\end{equation}
When solving the coupled dynamics between $c_Z(t)$ and $c_S(t)$, the contribution from the coupling between $c_Z(t)$ and $c_D(t)$ can be neglected, as it introduces only a quadratic dependence on $\beta_T\omega$ in the solution for $c_Z(t)$. Additionally, we note that 
$\sigma_{SZ}\approx0$, when $\kappa\approx1$.
Thus, considering the initial condition $c_Z(0)=0$, $c_S(0)={\frac{\sqrt{3}}{2}}$ and $\kappa\approx1$, leads to
\begin{equation}
\label{sol_ZS}
   \begin{split}
       &\langle I_Z\rangle(t)\propto c_Z(t)=\frac{\beta_T\omega_0}{\sqrt{2}}(e^{-t/T_1}-1)\\
       &\hspace{2.1cm}+\frac{\sqrt{3}}{2}\frac{\sigma_{ZS}}{T_1^{-1}-T_s^{-1}}(e^{-t/T_s}-e^{-t/T_1}),\\
       &\hspace{2cm}\langle\vec{I}_1\cdot\vec{I}_2\rangle(t)\propto c_S(t)={\frac{\sqrt{3}}{2}}e^{-t/T_s}.
   \end{split}
\end{equation}
As expected, our result coincides with that of references \cite{rodin2021,bengs2020}.
This example demonstrates that the correction term in the HTME
equation is essential to accurately describe the dynamics of a system initially in a non-thermal state. Indeed, using ARH-IME leads to no coupling and results in a single exponential decay for the magnetization within a time $T_1$.

\section{High-temperature approximation in operator form. Abragam's argument}
\label{Sec.V} 

This section reexamines the strategy used by Abragam to derive  Eq.(\ref{inhomog_eq}), highlighting the subtleties of this historical approach. We build on this approach to re-obtain our HTME, based solely on the high-temperature hypothesis, without resorting to weak order. The analysis elucidates some quantum microscopic aspects, complementing  Sec.\ref{Sec:3} by revealing fundamental features of the approximation.

We first examine how the high-temperature condition affects the operator structure of the full-quantum terms in Eq.(\ref{bornmarkov_rearranged}). Thus, we focus on the operator

\begin{equation}
\label{O_operator} 
{\cal O}\equiv \int_{-\infty}^{\infty}dt'\; [H_I(t'),\rho_B],
\end{equation}
where $H_I(t')= e^{i(H_S + H_B) t'} H_I\; e^{-i (H_S + H_B) t'}$. In Eq.(\ref{O_operator}), we defined  $t'\equiv t- \tau$ 
to facilitate the comparison with reference \cite{abragam61}. 

Operator ${\cal O}$ is crucial in defining the quantum characteristics of spin dynamics at microscopic time scales.
Using 
 the set $\{\ket{\alpha}\otimes\ket{f} \}$, where $\ket{\alpha},\ket{f}$ are eigenvectors of $H_S$ and $H_B$ respectively, as basis for the composite Hilbert space, a general matrix element can be unfolded as

\begin{equation}
\label{integral_elematriz} 
{\cal O}_{\alpha f\alpha' f'} =\int_{-\infty}^{\infty} dt' e^{it'(\alpha + f - \alpha'-f')} \langle \alpha f|  H_I  \;  | \alpha' f'\rangle\; 
\frac{e^{-\beta_T f'} - e^{-\beta_T f}}{Z}
,
\end{equation}
where the integral over $t'$
\begin{equation}
\label{delta}
\int_{-\infty}^{\infty} dt' e^{it'(\alpha + f - \alpha'-f')}= 2\pi \; \delta (\alpha + f - \alpha'-f')
\end{equation}
imposes  the condition on the eigenvalues of $H_S$ and $H_B$
\begin{equation}
\label{conser-energy}
\beta_T(\alpha' - \alpha) = \beta_T(f - f') \,,
\end{equation}
which implies a constraint that the corresponding eigenstates  must meet in the density matrix elements
 during the partial reduction process, and represents a kind of correlation between both systems through the spin-lattice interaction.
Thus, combining (\ref{conser-energy}) and  (\ref{integral_elematriz}), we obtain
\begin{equation}
\label{elematriz-conser}
\begin{split}
    {\cal O}_{\alpha f\alpha' f'} = \int_{-{\infty}}^{\infty} dt' &e^{it'(\alpha + f - \alpha'-f')}\\
    &\times\langle \alpha f|  H_I  \;  | \alpha' f'\rangle\; 
\frac{e^{-\beta_Tf' }}{Z}(1 - e^{-\beta_T(\alpha' - \alpha) } )\;.
\end{split}
\end{equation}
The factor $\beta_T f$ is not universally small, so the exponential $e^{-\beta_T f}$ in \eqref{elematriz-conser} cannot be 
truncated to its leading Taylor series term. However, when the condition $|\beta_T (\alpha - \alpha')| \ll 1$ is met, we may approximate $e^{-\beta_T(\alpha' - \alpha)} \simeq 1 - \beta_T(\alpha' - \alpha)$. This amounts to neglecting contributions of order $(\beta_T\omega)^2$ and higher in the spin dynamics.

  In this way, we assume that Eq. \eqref{elematriz-conser} can be approximated by its linear contribution,
\begin{equation}
\label{elematriz-aprox}
\begin{split}
    {\cal O}_{\alpha f\alpha' f'}^{HT}  \equiv \int_{-\infty}^{\infty} \;dt' &e^{it'(\alpha + f - \alpha'-f')} \\
    &\times\langle \alpha f|  H_I \;  | \alpha' f'\rangle\; 
\frac{1}{Z}e^{-\beta_T f' } \beta_T(\alpha' - \alpha).
\end{split}
\end{equation}
Thus, operator $\cal O$ can be formally expressed as follows: 
\begin{equation}
{\cal O }= {\cal O }_{HT} + \cal O',
\end{equation}
where $\cal O'$ comprises the operators related to the higher order terms in the expansion (\ref{elematriz-conser}). 
The high-temperature approximation then consists of retaining only 
${\cal O }_{HT}$ in the master equation (\ref{bornmarkov_rearranged}) and discarding the contributions associated with $\cal O'$.

The operator ${\cal O }_{HT}$, whose matrix elements are given by (\ref {elematriz-aprox}), can now be recast into 
\begin{equation}
\label{operador-aproximado}
{\cal O}_{HT}\equiv \int_{-\infty}^{\infty}dt'\; [H_I(t'),\beta_T H_S]\;\rho_B\;.
\end{equation}

Notice that the structure of ${\cal O }_{HT}$
is radically different from that of ${\cal O }$, since now the commutation corresponds to operators acting on the spin Hilbert space only, while the bath state is shifted out from the commutator.
We may also assume that the system equilibrium state $\rho_{Seq}$ can be approximated by 
\begin{equation}
\label{equilibrio_lineal}
\frac{1}{Z}\{\mathbb{I} - \beta_T H_S\} \simeq \rho_{Seq}=\frac{e^{-\beta_T H_S }}{\tr (e^{-\beta_T H_S }) }\; ,
\end{equation}
where $Z$ labels the number of degrees of freedom of the spin system. Then, replacing (\ref{operador-aproximado}) and   (\ref{equilibrio_lineal}) into the Born-Markov equation (\ref{bornmarkov_rearranged}), we get
\begin{equation}
\label{BM_HT}
\begin{split}
\frac{d\rho_S}{dt} = &-\frac{1}{2} \tr_B\left\{\int_{-{\infty}}^{\infty}dt' 
[H_I(t),[H_I(t'), \rho_S(t)   ]]\rho_B\right.\\
&-H_I(t) Z \rho_S(t)\int_{-{\infty}}^{\infty} dt'[H_I(t'),\rho_{S}^{eq}] \rho_B\\
&\left.+\int_{-{\infty}}^{\infty}dt'  [H_I(t'),\rho_{S}^{eq}]\rho_B \; Z\rho_S(t)H_I(t)\right\}.
\end{split}
\end{equation}
As anticipated, replacing the full operator ${\cal O}$
with its linearized form ${\cal O}_{HT}$ in 
(\ref{bornmarkov_rearranged}) amounts to neglecting higher-order contributions in $\beta H_S$. Crucially, this substitution also modifies the original structure of the full-quantum terms in Eq. (\ref{bornmarkov_rearranged}).

At this stage, a consistency analysis is required to verify that our approach: (i) correctly reproduces the finite-temperature equilibrium steady state, and (ii) enables accurate interpretation of the underlying physics.

First, we observe  that the transformation $\mathcal{O}\to\mathcal{O}_{HT}$ due to the high temperature hypothesis has consequences on the kind of spectral densities that can be compatible with this approximation.
Due to relation (\ref {conser-energy}), the Boltzmann factor in \eqref{elematriz-aprox} can be expressed as
\begin{equation}
\label{linear_boltz}
    e^{-\beta_T f'}= e^{-\beta_T f}e^{\beta_T(\alpha - \alpha')} = e^{-\beta_T f}(1 + \beta_T(\alpha - \alpha') + \cdots ).
\end{equation}
Then, under the high temperature regime,  \eqref{elematriz-aprox}  also satisfies
\begin{equation}
\label{elematriz-aprox_B}
\begin{split}
    {\cal O}_{\alpha f\alpha' f'}^{HT}  \equiv \int_{-\infty}^{\infty} \;dt'& e^{it'(\alpha + f - \alpha'-f')}\\
    &\times\langle \alpha f|  H_I \;  | \alpha' f'\rangle\; 
\frac{e^{-\beta_T f } }{Z}\beta_T(\alpha' - \alpha).
\end{split}
\end{equation}
This shows that in this limit,  the matrix elements of operator ${\cal O}_{HT}$
become insensitive to replacing the bath eigenvalue in the Boltzmann factor, i.e.
$e^{\beta_T f'} \rightarrow e^{\beta_T f}$.
Thus, in terms of operators, the high-temperature hypothesis implies assuming the  equality
\begin{equation}
\label{operadores-similares}
{\cal O}_{HT}=\int_{-\infty}^{\infty}dt' \; [H_I(t'),\beta_T H_S]\;\rho_B = \int_{-\infty}^{\infty}dt'\;\rho_B \; [H_I(t'),\beta_T H_S]
.
\end{equation}
We notice that this subtle argument 
implies equating operators that otherwise would  be different. Due to this equivalence, the third term of \eqref{BM_HT} can be changed as
\begin{equation}
\label{reemplazo}
\begin{split}
&\hspace{2.3cm}C\longrightarrow D\\
    &C \equiv \int_{-{\infty}}^{\infty}dt' \;  \tr_B\{[H_I(t'),\rho_{S}^{eq}] \rho_S  \rho_B H_I(t) \}\\
    & D \equiv \int_{-{\infty}}^{\infty}dt' \;\tr_B\{[H_I(t'),\rho_{S}^{eq}] \rho_S H_I(t) \rho_B\}\,.
\end{split}
\end{equation} 
If substitution (\ref{reemplazo}) were performed, all the terms in (\ref{BM_HT}) could be viewed as operators averaged over the lattice degrees of freedom. 

Using the explicit expressions in (\ref{hamiltonianos}) for the interaction Hamiltonian and discarding non-secular terms, we obtain  
\begin{equation}
\label{A}
\begin{split}
    & C= \sum_{\alpha \beta,\omega}   [A_{\beta}(\omega),\rho_{Seq}]\rho_S A^\dagger_{\alpha}(\omega) \; {\cal J}_{\alpha \beta}(\omega),\\
    & D= \sum_{\alpha \beta,\omega}   [A_{\beta}(\omega),\rho_{Seq}]\rho_S A^\dagger_{\alpha}(\omega) \; {\cal K}_{\alpha \beta}(\omega)
\end{split}
\end{equation}
where we assumed stationarity of the lattice and employed the spectral density definitions
\eqref{denpec_defin} and \eqref{Kdenpec_defin}.
The equivalence of $C$ and $D$ means that  the high-temperature spectral densities  satisfy the symmetry relationships  
${\cal J}_{\alpha\beta}(\omega) = {\cal K}_{\alpha\beta}(\omega)$ or
${\cal J}_{\alpha\beta}(-\omega) = {\cal J}_{\beta \alpha}(\omega)$ instead of \eqref{bce_detallado} and \eqref{bce_detallado_b}.
This symmetry property of the spectral densities is also evident
from Eqs. (\ref{denpec_defin_app}-\ref{Kdenpec_defin_app})
where it can be seen that equating $e^{\beta_T f}=e^{\beta_T f'}$
renders all spectral density definitions identical. Notice that this characteristic is unique to the linear approximation in $\beta_T\omega$, as it disappears when including quadratic terms in the expansion of (\ref{elematriz-aprox_B}).

This feature is a symptom of the loss of quantumness caused by the high temperature regime.
Physically, the symmetry reflects a fundamental change in description at the microscopic level in consistency with the linear approximation of the master equation, in which the spectral densities are transformed but continue to have quantum character. The dependency on $\beta_T\omega$ remains, but now based on a weaker quantum interrelationship or correlation
between eigenlevels and eigenstates of the Hamiltonians involved in the partial reduction process.
If the nature of the system-environment interaction is of very low quantumness, such that a stochastic approximation is adequate, then a semiclassical hypothesis can be considered an extreme limit of the high-temperature approximation.

At this point, it is worth mentioning that Hubbard introduced symmetric spectral densities in his derivation of the ARH-IME \cite{hubbard61}. In short, the author proposed to define \textit{symmetric} bath correlation functions in terms of the quantum correlation functions. In this way, he obtained a symmetric version of the quantum spectral densities and
expressed the master equation in terms of these functions.
Then, after applying the high-temperature limit, he arrives at the inhomogeneous master equation. A similar expression of the master equation in terms of symmetric spectral densities can be found in \cite{rodin2021}. These authors argue that writing the master equation in terms of symmetrized spectral densities facilitates the transition to a semiclassical approximation.
However, these works did not examine the explicit transformation of spectral densities under the high-temperature limit, a gap our present analysis addresses.

We now apply the formalism used in Section \ref{Sec:3} to express Eq.\eqref{BM_HT}
in terms of eigen-operators and apply the secular approximation. Thus, after replacing the expressions given in  Eq.(\ref{hamiltonianos}) for the interaction Hamiltonian and using the form (\ref{desa-rho-llo}) for the reduced density operator, we obtain
\begin{equation}
\begin{split}
\label{c-terms_linear}
\frac{d\rho_S}{dt}=
- \frac{1}{2} &\sum_{\alpha,\beta,\omega}{\cal J}_{\alpha,\beta}(\omega)\{[A^\dagger_\alpha(\omega), [A_\beta(\omega), \rho_S(t) ]\\
&+ \beta_T \omega[A_\beta(\omega), \rho_S(t)] 
A^\dagger_\alpha(\omega) 
 - [A_\alpha^\dagger (\omega), [A_\beta (\omega),\rho_{Seq}    ]  \,  ]\\
&
- A_\alpha^\dagger (\omega) \Delta \rho_S  [A_\beta (\omega),\rho_{Seq}    ] +  [A_\beta (\omega),\rho_{Seq}    ] \Delta \rho_S A_\alpha^\dagger (\omega) \}.
\end{split}
\end{equation}
The first two terms corresponds to the linear-in-$\beta_T\omega$ approximation of Eq. (\ref{c-terms}), 
the third and the two last terms come from replacing $\rho_S Z$ by the identity and $\Delta\rho_S$ respectively
 in Eq.(\ref{BM_HT}). Finally, separating the sums over positive and negative frequencies and using (\ref{conmut_Hs}), (\ref{rho_eq}) and (\ref{bce_detallado_b}) after some  algebra Eq.(\ref{c-terms_linear}) can be written as
\begin{equation}
\begin{split}
\label{abrag_lin}
\Dot{\rho}_S=&  -\frac {1}{2}\sum_{\alpha,\beta,\omega} { J}_{\alpha,\beta}(\omega)\Big\{ [A^\dagger_\alpha(\omega), [A_\beta(\omega),  { \rho}(t,\beta_T\omega) - \rho_S^{eq} ]]\\
&
\hspace{1cm}-[{\Delta \rho_S} (t,\beta_T\omega)      
[\rho_S^{eq}, A_\beta(\omega)],A^\dagger_\alpha(\omega) ]\Big\}\\
&-\frac{1}{2}\sum_{\alpha\beta}\sum_{\omega>0}J_{\alpha\beta}(\omega)[[\rho_S^{eq},A_{\alpha}(\omega)],[A_{\beta
        }^{\dagger}(\omega),\Delta\rho_S]]).
\end{split}
\end{equation}
This result coincides with Eq.(\ref{ecmaes-eigen_aprox_b}) and represents the Markovian spin dynamics in the high-temperature limit, which is made up of the ARH-IME term and a contribution which can be relevant in cases of non-thermal states far from equilibrium.
The last term in (\ref{abrag_lin})
arises from considering the linear contribution of the detailed balance relation (\ref {bce_detallado_b}) which is made explicit when expressing the Born-Markov equation in terms of eigen-operators, as was done in Sec. \ref{Sec.II}.
Notice that this contribution does not arise from strictly applying the formalism used by Abragam since he did not use the spectral expansion of the interaction Hamiltonian prior to applying the high temperature hypothesis, as we did above. 

\subsection{Weak order condition. ARH-IME}
The previous analysis reveals the dramatic changes in the structure of the master equation as the high-temperature limit is considered. Remarkably, this limit allows us to commute bath operators up to first order in $\beta_T\omega$. Consequently, the spectral densities become symmetric, which validates our reasoning in Sec. \ref{Sec:3}. 
Making use of these features of the approximation, we can rewrite Eq.(\ref{BM_HT}) as follows:

\begin{equation}
\label{BM_HT-permutada}
\begin{split}
\frac{d\rho_S}{dt} = -\frac{1}{2} & \tr_B\left\{\int_{-{\infty}}^{\infty}dt'
[H_I(t),[H_I(t'), \rho_S(t)   ]]\rho_B\right.\\
&+\int_{-{\infty}}^{\infty} dt'\;H_I(t) Z \rho_S(t) \; [H_I(t'),\rho_{Seq}] \rho_B\\
&\left. -  \int_{-{\infty}}^{\infty}dt'\ [H_I(t'),\rho_{Seq}] Z \rho_S(t) H_I(t)\; \rho_B \right\},
\end{split}
\end{equation}
where in the last term operators $H_I(t)$ and $\rho_B$ were permuted. 
Finally, expressing $\rho_S$
as in (\ref{desa-rho-llo})
and adding the weak-order condition, i.e., $\rho_S\approx\frac{I}{Z}$ in the second and third terms, we get 
\begin{equation}
\label{BM_HT-Abragam}
\left(\frac{d\rho_S}{dt} \right)_{AR} = -\frac{1}{2}  \tr_B \int_{-{\infty}}^{\infty}dt'
[H_I(t),[H_I(t'), \rho_S(t) - \rho_{Seq}   ]]\rho_B.
\end{equation}
We recall that Abragam arrived at the famous \textit{inhomogeneous} NMR master equation in this way (see section VIII-D of \cite{abragam61}). So, in conclusion, we can say that Abragam's derivation of the master equation is the linearization of the Lindblad or Born-Markov equations but restricted to weak-order initial states only. 

\section{\label{Sec.Concl} Discussion And Analysis}

In this work, we investigated  the transformation of the quantum Markovian master equation in the high-temperature limit. 
In first place, we derived  Lindblad's equation from the Born-Markov equation, exposing the total dependency on the Boltzmann factor originated from the detailed-balance relations. At this point, we apply the high-temperature condition to keep only those contributions that are linear in $\beta_T\omega$.
The strategy utilized in this work allows us to rigorously identify the Abragam-Redfield-Hubbard
equation as one term of the total contribution, and reveal the emergence of a remainder term which contributes in special cases, such as when the
spin system is prepared in a non-thermal initial state, e.g. a singlet state.
This equation correctly describes the thermalization of the system at finite temperatures {due to the quantum dynamical system-bath interaction at high temperatures, for any type of initial preparation of the spin system.

From a physical perspective, the high-temperature approximation is justified in NMR systems under common conditions. The energy quanta of nuclear spin transitions are typically much smaller than the thermal energy. Thus, a linear approximation of the  series expansion of the master equation in powers of
$\beta_T\omega$, in the absence of collective processes capable of compensating for the smallness of the powers of $\beta_T\omega$, seems reasonable. The high-temperature master equation exhibits key modifications on both characteristic timescales: the structure of the equation (coarse-grained scale) changes when the high-temperature limit is adopted, and consistently the spectral densities (microscopic scale) become symmetric, reflecting a fundamental change in the relationship between the interaction Hamiltonian $H_I(t)$ and the bath density matrix $\rho_B$. This indicates diminished quantum effects of the system-environment interaction. 
The entire change ensures correct thermal equilibrium at finite temperatures.

To illustrate these findings, we analyze  two examples of very distinct characteristics, 
  namely, an isolated spin--$\frac{1}{2}$ coupled to a bosonic bath, and a spin--$\frac{1}{2}$ pair subjected to a random magnetic field and prepared non-thermally in a singlet state. In the first case, the full-quantum approach is accessible, allowing us to capture the symmetrization of the spectral densities as the high temperature limit is taken, with the consequence that at this limit the mechanism responsible of the spontaneous emission in the TLS is suppressed, giving rise to microscopic reversibility represented by the symmetric spectral densities. 
  
  Furthermore, in this simple academic example, we note that the magnetization dynamics (spin polarization) is described by the Bloch equations for all types of initial preparation—including non-thermal states far from equilibrium—as expected. This demonstrates the validity of the ARH-IME even in these extreme cases. However, the limitations of the HTME become apparent when describing relaxation due to vacuum fluctuations. 
  
  This simple academic example serves to highlight some key points: first, that the ARH-IME remains valid for non-thermal initial states far from equilibrium, as the dynamic of the magnetization is described (as expected) by the Bloch equations for any preparation; and second, that the HTME exhibits clear limitations when describing relaxation due to vacuum fluctuations. In this sense, and following Redfield \cite{redfield_1965}, one way to account for this low-temperature behaviour would be weighting the high-temperature version of the spectral density (\ref{eq:_optic_spectral_density}) with the factor $e^{\frac{1}{2}\beta_T\omega}$, and use it in Eq.(\ref{eq:Eq_Motion_Optic}) . However, as shown in appendix \ref{AppD}, in the limit $T\to0$ this lead to a zero time derivative for the state of the system and thus to no decay of the polarization. Therefore, weighting the spectral densities in this way, which resembles the approach of reference \cite{bengs2020}, is not enough to account for the low-temperature effects. Accordingly, it is reasonable to encounter this kind of discrepancy at low temperatures whenever an approach based on using classical spectral densities is adopted. 

In the other example, the initial state is a non-thermal singlet state (maximally entangled), and the spin system is coupled to a fluctuating random field, with the particularity that the fluctuations in both spins are correlated. This example was used in references \cite{bengs2020,rodin2021} to discuss the singlet-triplet conversion, in the framework of the study of longlived states.
Unlike the ARH-IME equation, our generalized high-temperature master equation captures the coupling between Zeeman and singlet orders, essential for describing singlet-triplet conversion in long-lived states. This case highlights the limitations of the ARH-IME framework when dealing with correlated baths and non-thermal high-order preparations.

\section{Summary}

Our results could be summarized as follows: 

\textbf{1)} {We found that in the high-temperature limit, the Born-Markov and, consequently, the Lindblad equation, converge to the 
ARH-IME when weak order is assumed.  Therefore, the inhomogeneity is not simply an ad hoc factor introduced to predict the correct steady state, but its emergence is a rigorous result.
Our procedure clarifies the origin of the inhomogeneity in the weak-order high-temperature equation that was introduced by Redfield, Abragam and Hubbard during the early days of relaxation theory.   }

\textbf{2)} We observe that linearizing the Lindblad equation yields the ARH-IME, modified by a term that may become relevant outside the weak-order regime. The significance of this new contribution became evident when analyzing the problem of a spin-1/2 pair interacting with a random magnetic field. There, it was demonstrated that the ARH-IME equation alone does not suffice to describe the system's dynamics when the spin system is prepared in a non-thermal state far from equilibrium and coupled to a correlated bath.
Therefore, including this new contribution becomes essential for accurately modelling this system.
This analysis also shows that, regardless of the initial preparation of the system, the Lindblad equation converges asymptotically to the ARH-IME when the system approaches thermal equilibrium.

\textbf{3)} 
The procedure carried out in section \ref{Sec:3} to obtain the HTME 
reveals that retaining only linear terms in the expansion leads to symmetrical spectral densities, i.e. $J_{\alpha\beta}(\omega)=J_{\beta\alpha}(-\omega)$.
This symmetry, confirmed in Sec. \ref{Sec.V} using Abragam's formalism, reflects a fundamental suppression of quantum effects in the spin-lattice interactions. Crucially, however, this does not justify a priori classical treatment of the environment--as illustrated in Sec. \ref{Sec:4} for a spin-1/2 system coupled to a bosonic bath, where spectral density symmetrization occurs within a fully quantum framework. Nevertheless, this symmetry helps to explain the success of empirical semi-classical models in the high-temperature regime, where stochastic approximations can effectively mimic residual quantum behavior.

\section{Conclusion}

In this work, we derived a quantum Markovian master equation valid under high-temperature conditions, describing spin-lattice relaxation under the sole additional assumption of the linear expansion’s validity. This assumption leads to profound changes in both the operator form of the Markovian master equation and the microscopic quantum properties of the system-environment interaction it can represent.
Critically, these modifications symmetrize the spectral densities. Physically, this restricts the equation’s applicability to high-temperature regimes where quantum effects in the system-environment interaction are weak. Spin-lattice relaxation in liquid NMR could exemplify this scenario.
Our results show that the ARH-IME—a cornerstone of NMR relaxation theory—is a specific case of the broader Born-Markov and Lindblad formalisms under high-temperature and weak-order conditions. When the latter condition fails, the equation naturally extends, within the open quantum systems framework, to the HTME to account for higher-order or spin-correlation effects. 

We anticipate that this work will stimulate further discussion on the applicability of the inhomogeneous ARH-IME and the nature of the high-temperature quantum master equation in open systems. 
Future research could overcome the limitations imposed by linearizing, and explore spin relaxation in regimes where the system-environment interaction exhibits stronger quantum effects.

\appendix
\section{Derivation of Eq. (5)}
\label{AppendixA}
We aim in this appendix to include all the calculations omitted in the main body of this work. We begin by proving eq. (\ref{bornmarkov_rearranged}) using the Hamiltonian's eigenoperator representation. 

We begin by expanding the double-commutator operator in Eq.(5)
\begin{equation}
\label{eq:A1}
    \begin{split}
        &[H_I(t),[H_I(t'),\rho_S(t)   ]]\rho_B=H_I(t)H_I(t')\rho_S(t)\rho_B\\
        &-H_I(t){\rho}_S(t)H_I(t')\rho_B-H_I(t'){\rho}_S(t)H_I(t)\rho_B\\
        &+\rho_S(t)H_I(t')H_I(t)\rho_B
    \end{split}
\end{equation}
We do the same with the double commutator in Eq.(\ref{bornmarkov})
\begin{equation}
\label{eq:A.2}
    \begin{split}
&[H_I(t),[H_I(t'),\rho_S(t)\rho_B]]=H_I(t)H_I(t')\rho_S(t)\rho_B\\
        &-H_I(t){\rho}_S(t)\rho_BH_I(t')-H_I(t'){\rho}_S(t)\rho_BH_I(t)\\
        &+\rho_S(t)\rho_BH_I(t')H_I(t)
    \end{split}
\end{equation}
Now we add and subtract the operator $H_I(t){\rho}_S(t)H_I(t')\rho_B+H_I(t'){\rho}_S(t)H_I(t)\rho_B$ to Eq.(\ref{eq:A.2}), then
\begin{equation}
    \label{eq:A.3}
    \begin{split}
&[H_I(t),[H_I(t'),\rho_S(t)\rho_B]]=H_I(t)H_I(t')\rho_S(t)\rho_B\\
        &-H_I(t){\rho}_S(t)\rho_BH_I(t')-H_I(t'){\rho}_S(t)\rho_BH_I(t)\\
        &+\rho_S(t)\rho_BH_I(t')H_I(t)+H_I(t){\rho}_S(t)H_I(t')\rho_B+H_I(t'){\rho}_S(t)H_I(t)\rho_B\\
        &-H_I(t){\rho}_S(t)H_I(t')\rho_B-H_I(t'){\rho}_S(t)H_I(t)\rho_B
    \end{split}
\end{equation}
We split Eq.(\ref{eq:A.3}) into three parts
\begin{equation}
\label{eq:A4}
    \begin{split}
        &H_I(t)H_I(t')\rho_S(t)\rho_B-H_I(t){\rho}_S(t)H_I(t')\rho_B\\
        &-H_I(t'){\rho}_S(t)H_I(t)\rho_B
        +\rho_S(t)\rho_BH_I(t')H_I(t)
    \end{split}
\end{equation}
Notice that in the last term, we can shift $\rho_B$ to right under partial trace, that is,
\begin{equation*}
\tr_B(\rho_S(t)\rho_BH_I(t')H_I(t))=\tr_B(\rho_S(t)H_I(t')H_I(t)\rho_B),
\end{equation*}
then, under the partial trace, Eqs (\ref{eq:A1}) and (\ref{eq:A4}) are equal.

The remaining contributions are
\begin{equation}
\label{eq:A5}
    H_I(t){\rho}_S(t)H_I(t')\rho_B-H_I(t){\rho}_S(t)\rho_BH_I(t')=H_I(t)\rho_S[H_I(t'),\rho_B]
\end{equation}
and
\begin{equation}
\label{eq:A6}
    H_I(t'){\rho}_S(t)H_I(t)\rho_B-H_I(t'){\rho}_S(t)\rho_BH_I(t)
\end{equation}
In the first term in Eq.(\ref{eq:A6}) we shift $\rho_B$ to the left under partial trace, that is
\begin{equation}
    \begin{split}
        &\tr_B(H_I(t'){\rho}_S(t)H_I(t)\rho_B-H_I(t'){\rho}_S(t)\rho_BH_I(t))\\
        &=\tr_B(\rho_BH_I(t'){\rho}_S(t)H_I(t)-H_I(t'){\rho}_S(t)\rho_BH_I(t))\\
        &=-\tr_B([H_I(t'),\rho_B]H_I(t)\rho_S)
    \end{split}
\end{equation}
Then, by joining Eqs. (\ref{eq:A4}), (\ref{eq:A5}) and (\ref{eq:A6}), we get that 
\begin{equation}
   \begin{split}
       &\tr_B([H_I(t),[H_I(t'),\rho_S(t)\rho_B]])
       \\&
       = \tr_B([H_I(t),[H_I(t'),\rho_S(t)   ]]\rho_B+H_I(t)\rho_S[H_I(t'),\rho_B]\\
       &-H_I(t'),\rho_B]H_I(t)\rho_S)
   \end{split}
\end{equation}

\section{Derivation of Eq. (9.a) and (9.b)}
\label{AppendixB}

In what follows, we derive equations (\ref{q-terms}) and (\ref{c-terms}). In all the expressions the secular approximation is applied.
\begin{equation}
\begin{array}{ll}
    &\int_{\mathbb{R}}\tr_B([H_I(t),[H_I(t'),\hat\rho_S(t)]]\rho_B)dt'\\
&=\sum_{\alpha\beta}\sum_{\omega}\int_{\mathbb{R}}d\tau e^{i\omega\tau}[(A^{\dagger}_{\alpha}A_{\beta}\rho_S-{A^{\dagger}_{\alpha}\rho_SA_{\beta})\langle B_{\alpha}(0)B_{\beta}(\tau)\rangle}\\
    &\\
        &\hspace{30mm}+(\rho_SA_{\beta}A_{\alpha}^{\dagger}-A_{\beta}\rho_SA_{\alpha}^{\dagger})\langle B_{\beta}(\tau)B_{\alpha}(0)\rangle]\\
        &\\
        &=\sum_{\alpha\beta}\sum_{\omega}\mathcal{J}_{\alpha\beta}(\omega)(A^{\dagger}_{\alpha}A_{\beta}\rho_S-A^{\dagger}_{\alpha}\rho_SA_{\beta})\\
        &\hspace{1.3cm}+\mathcal{K}_{\beta\alpha}(\omega)(\rho_SA_{\beta}A_{\alpha}^{\dagger}-A_{\beta}\rho_SA_{\alpha}^{\dagger})
\end{array}
\end{equation}
In the last line of eq. (\ref{eq:A6}) we have introduced the following quantities
\begin{equation}
   \begin{split}
       &\mathcal{J}_{\alpha\beta}(\omega)=\int_{\mathbb{R}}d\tau e^{i\omega\tau}\langle B_{\alpha}(0)B_{\beta}(\tau)\rangle,\\
       &\mathcal{K}_{\alpha\beta}(\omega)=\int_{\mathbb{R}}d\tau e^{i\omega\tau}\langle B_{\beta}(\tau)B_{\alpha}(0)\rangle,
   \end{split}
\end{equation}
from which it can be proven the identity: 
\begin{equation}
\label{eq:identity}
    \mathcal{K}_{\alpha\beta}(\omega)=\mathcal{J}_{\beta\alpha}(-\omega)=e^{-\beta_T\omega}\mathcal{J}_{\alpha\beta}(\omega).
\end{equation}
Using Eq. (\ref{eq:identity}), we get Eq.(\ref{c-terms}) as follows
\begin{equation}
\label{B4}
  \begin{split}
       &\int_{\mathbb{R}}\tr_B([H_I(t),[H_I(t'),\hat\rho_S(t)]]\rho_B)dt'\\
       &=\sum_{\alpha\beta}\sum_{\omega}\mathcal{J}_{\alpha\beta}(\omega)(A^{\dagger}_{\alpha}A_{\beta}\rho_S-A^{\dagger}_{\alpha}\rho_SA_{\beta}+e^{-\beta_T\omega}(\rho_SA_{\beta}A_{\alpha}^{\dagger}-A_{\beta}\rho_SA_{\alpha}^{\dagger}))\\        &=\sum_{\alpha\beta}\sum_{\omega}\mathcal{J}_{\alpha\beta}(\omega)[A^{\dagger}_{\alpha}A_{\beta}\rho_S-A^{\dagger}_{\alpha}\rho_SA_{\beta}+\rho_SA_{\beta}A_{\alpha}^{\dagger}-A_{\beta}\rho_SA_{\alpha}^{\dagger}\\
        &\hspace{26mm}+(e^{-\beta_T\omega}-1)(\rho_SA_{\beta}A_{\alpha}^{\dagger}-A_{\beta}\rho_SA_{\alpha}^{\dagger})]\\
    &=\sum_{\alpha\beta}\sum_{\omega}\mathcal{J}_{\alpha\beta}(\omega)([A_{\alpha}^{\dagger},[A_{\beta},\rho_S]]+(e^{-\beta_T\omega}-1)[\rho_S,A_{\beta}]A_{\alpha}^{\dagger}\\
    &=\sum_{\alpha\beta}\sum_{\omega}\mathcal{J}_{\alpha\beta}(\omega)([A_{\alpha}^{\dagger},[A_{\beta},\rho_S]]+(1-e^{-\beta_T\omega})[A_{\beta},\rho_S]A_{\alpha}^{\dagger}.
  \end{split}
\end{equation}
The same procedure yields to Eq. (\ref{q-terms})
\begin{equation}
\label{eq:A.10}
   \begin{split}
           &\int_{\mathbb{R}}\tr_B(H_I(t) \rho_S(t)[H_I(t'),\rho_B]-[H_I(t'),\rho_B] \rho_S(t) 
H_I(t))dt'\\
&=\sum_{\alpha\beta}\sum_{\omega}\int_{\mathbb{R}}d\tau e^{i\omega\tau}[(A_{\alpha}^{\dagger}\rho_SA_{\beta}-A_{\beta}\rho_SA_{\alpha}^{\dagger})\langle B_{\alpha}(0)B_{\beta}(\tau)\rangle\\
&\hspace{2.3cm} +(A_{\beta}\rho_SA_{\alpha}^{\dagger}-A_{\alpha}^{\dagger}\rho_SA_{\beta})\langle B_{\beta}(\tau)B_{\alpha}(0)\rangle]\\
&=\sum_{\alpha\beta}\sum_{\omega}\mathcal{J}_{\alpha\beta}(\omega)(1-e^{-\beta_T\omega})(A_{\alpha}^{\dagger}\rho_SA_{\beta}-A_{\beta}\rho_SA_{\alpha}^{\dagger}).
   \end{split}
\end{equation}
Adding and subtracting  $A_{\alpha}^{\dagger}A_{\beta}\rho_S$, we get
\begin{equation}
\label{eq:A.11}
   \begin{split}
&\sum_{\alpha\beta}\sum_{\omega}\mathcal{J}_{\alpha\beta}(\omega)(1-e^{-\beta_T\omega})(A_{\alpha}^{\dagger}\rho_SA_{\beta}-A_{\beta}\rho_SA_{\alpha}^{\dagger}\\
&+A_{\alpha}^{\dagger}A_{\beta}\rho_S-A_{\alpha}^{\dagger}A_{\beta}\rho_S)\\
    &=\sum_{\alpha\beta}\sum_{\omega}\mathcal{J}_{\alpha\beta}(\omega)(1-e^{-\beta_T\omega})([A_{\alpha}^{\dagger},A_{\beta}\rho_S]+A_{\alpha}^{\dagger}[\rho_S,A_{\beta}])\\
    &=\sum_{\alpha\beta}\sum_{\omega}\mathcal{J}_{\alpha\beta}(\omega)(1-e^{-\beta_T\omega})(-[A_{\beta}\rho_S,A_{\alpha}^{\dagger}]+A_{\alpha}^{\dagger}[\rho_S,A_{\beta}])
   \end{split}
\end{equation}

\section{Spectral density detailed balance relationships}
\label{AppendixC}
\begin{equation}
\label{denpec_defin_apend}
  { \cal J}_{ \alpha \beta} (\omega)\equiv \int_{-\infty}^{\infty} \;d\tau \;
e^{i\omega \tau}  \;\tr\{ B_{\alpha}
B_{\beta}(-\tau)  \rho_B\}
\end{equation}
where $B_{\beta}(-\tau) = e^{-iH_B \tau} B_{\beta}\; e^{i H_B \tau}.$
\begin{equation}
\label{Kdenpec_defin_apend}
{ \cal K}_{ \alpha \beta} (\omega)\equiv \int_{-\infty}^{\infty} \;d\tau \;
e^{i\omega \tau}  \;\tr\{ 
B_{\beta}(-\tau) B_{\alpha} \rho_B       \}
\end{equation}
and
\begin{equation}
{ \cal J}_{ \beta \alpha} (\omega)\equiv \int_{-\infty}^{\infty} \;d\tau \;
e^{i\omega \tau}  \;\tr\{ B_{\beta}
B_{\alpha}(-\tau)  \rho_B       \}
\end{equation}

\begin{equation}
\label{Tr_BA}
\begin{split}
    \tr_B \{ B_{\beta}(-s) \rho_B B_{\alpha} \}&= \tr_B \{ B_{\alpha} B_{\beta}(-s) \rho_B  \}\\
    &=
\sum_{ff'} \langle f|B_{\beta}|f'\rangle \langle f'| B_{\alpha}|   f  \rangle \; \rho_{f'} \; e^{-i s (f-f')},
\end{split}
\end{equation}

\begin{equation}
\label{Tr_BB}
\tr_B \{ B_{\beta}(-s) B_{\alpha}  \rho_B  \}= 
\sum_{ff'} \langle f|B_{\beta}|f'\rangle \langle f'| B_{\alpha}|   f  \rangle \; \rho_f\; e^{-i s (f-f')}\;,
\end{equation}

According to the definitions of the spectral densities given in equations (\ref{denpec_defin_apend}) and (\ref{Kdenpec_defin_apend}), after making  the time correlation functions explicit, we obtain the following expressions

\begin{equation}
\label{denpec_defin_app}
{ \cal J}_{ \alpha \beta} (\omega)= 
\sum_{ff'} \;\delta(\omega - f' + f) \langle f| B_{\alpha}|   f'  \rangle 
\langle f'|B_{\beta}|f\rangle \; \rho_f\;
\end{equation}

\begin{equation}
{ \cal J}_{ \beta \alpha} (-\omega)=
\sum_{ff'} \;\delta(\omega - f' + f)\langle f|B_{\alpha}|f'\rangle \langle f'| B_{\beta}|   f  \rangle \; \rho_{f'}\;
   \end{equation}

\begin{equation}
\label{Kdenpec_defin_app}
{ \cal K}_{ \alpha \beta} (\omega)=
\sum_{ff'} \;\delta(\omega - f' + f)\langle f|B_{\alpha}|f'\rangle \langle f'| B_{\beta}|   f  \rangle \; \rho_{f'}\;.
\end{equation}
Thus, after applying the conditions imposed by the delta functions on the eigenvalues of the environment Hamiltonian, the relation (\ref{bce_detallado})
follows. 

\section{Low-Temperature limit}
\label{AppD}
To show the unsuitableness of Redfield weighting process, we apply it to the spin-boson model developed in sec. \ref{Sec:3}. To that end, we take the high-temperature spectral densities derived from Eq.(\ref{eq:_optic_spectral_density}), that is, $\mathcal{J}(\omega)=\frac{4\omega^3}{3\hbar c^3}(1+N(\omega))\approx \frac{4\omega^3}{3\hbar c^3}N(\omega)=\mathcal{J}(-\omega)$, weight them with the factor $e^{\frac{1}{2}\beta_T\omega}$ and use them in  Eq.(\ref{eq:Eq_Motion_Optic}). This gives 
\begin{equation}
\label{eq:Eq_Motion_Optic_L&B}
  \begin{split}
       \Dot{\rho}&=\gamma_0N(\omega)e^{-\frac{1}{2}\beta_T\omega}\left(\sigma_-\rho_S\sigma_+-\frac{1}{2}\sigma_+\sigma_-\rho_S-\frac{1}{2}\rho_S\sigma_+\sigma_-\right)\\
       &+\gamma_0N(\omega)e^{\frac{1}{2}\beta_T\omega}\left(\sigma_+\rho_S\sigma_--\frac{1}{2}\sigma_-\sigma_+\rho_S-\frac{1}{2}\rho_S\sigma_-\sigma_+\right).
  \end{split}
\end{equation}
Taking the limit $\beta_T\to\infty$, we get $N(\omega)=\frac{1}{e^{\beta_T\omega}-1}\approx e^{-\beta_T\omega}$, then $N(\omega)^{\pm\frac{1}{2}\beta_T\omega}\to 0$, and consequently, there is no decay of the expectation values $\langle\sigma_i\rangle$. 

However, taking the same limit in Eq.(\ref{eq:35}) gives
\begin{equation}
    \langle\Dot{\sigma}_{z}\rangle(t)=-\gamma_0\left(\langle\sigma_z\rangle(t)-1\right),\quad\langle\Dot{\sigma}_{\pm}\rangle=-\frac{1}{2}\gamma_0\langle\sigma_{\pm}\rangle(t).
\end{equation}

\section*{DATA AVAILABILITY}
The data that support the findings of this study are available within the article.

\section*{Acknowledgments}

This work was supported by SECYT, Universidad Nacional de Córdoba. J.A.T. thanks CONICET for financial support.

\section*{References}
\bibliographystyle{aipnum4-1}
\bibliography{References}

\begin{thebibliography}{36}
\bibitem [{foo()}]{footnote3}%
  \BibitemOpen
  \href@noop {} {\enquote {\bibinfo {title} {Comment on the validity of the approximation},}\ }\bibinfo {note} {This approximation could become invalid if some other phenomenon occurs that could compensate for the smallness of the thermal factor. In reference \cite{warren93}, it is shown that in some cases terms of higher order than linear in the expansion may be relevant for the dynamics. In that work, the observation of intermolecular MQC in bulk water and large molecules in solution was explained by discarding the usual approximation of the thermal equilibrium state together with the inclusion of the dipole-dipole interaction between spins separated by very large distances, considering all spins in the sample as a single quantum system. Although the article does not investigate relaxation dynamics, it serves as a reminder that discarding higher-order terms is also an alternative solution.}\BibitemShut {Stop}%
\end{thebibliography}

\end{document}